\documentclass[letterpaper,journal]{IEEEtran}
\usepackage{amsmath,amsfonts}
\usepackage{algorithmic}
\usepackage{algorithm}
\usepackage{array}
\usepackage[caption=false,font=normalsize,labelfont=rm,textfont=rm]{subfig}
\usepackage{textcomp}
\usepackage{stfloats}
\usepackage{url}
\usepackage{verbatim}
\usepackage{graphicx}
\usepackage{cite}
\usepackage{multirow}
\usepackage{xcolor}
\usepackage{gensymb}
\usepackage{amssymb}
\colorlet{BLUE}{blue}
\begin{document}

\title{A Compact Cross-Structured Dynamic Antenna for Reconfigurable Directional Modulation}

\author{Sheng Huang,~\IEEEmembership{Member,~IEEE}, Jacob R. Randall,~\IEEEmembership{Student Member,~IEEE}, Cory Hilton,~\IEEEmembership{Student Member,~IEEE}, Jeffrey A. Nanzer, \IEEEmembership{Senior Member,~IEEE}

\thanks{Manuscript received June 13, 2025. \textit{(Corresponding author: Jeffrey A. Nanzer.)}
} 
\thanks{The authors are with the Department of Electrical and Computer Engineering, Michigan State University, East Lansing, MI 48824 USA (e-mail:
	huang287@msu.edu; randa130@msu.edu; hiltonc2@msu.edu; nanzer@msu.edu).}
}
\markboth{IEEE Transactions on Antennas and Propagation,~Vol.~XX, No.~XX, 2026}%
{Huang \MakeLowercase{\textit{et al.}}: Cross-Structured Dynamic Antenna for Reconfigurable Information Beams}

\IEEEpubid{}

\maketitle

\begin{abstract}
A compact cross-structured dynamic antenna is presented for antenna-level physical-layer security using reconfigurable information-beam control rather than conventional radiation beam steering. The antenna consists of four printed meander-line monopoles arranged in a planar cross structure and driven by a switching network that realizes two complementary excitation states within each selected dynamic mode. By switching between opposite or diagonal port groups, the proposed aperture introduces apparent two-dimensional phase center displacement and supports four dynamic modes corresponding to information-beam directions of $\varphi=0^\circ$, $45^\circ$, $90^\circ$, and $135^\circ$. An average--differential array factor formulation is developed to show that the average radiation component preserves broad omnidirectional coverage, while the odd-symmetric differential component introduces angle-dependent magnitude and phase distortion that determines where the transmitted constellation remains recoverable. Thus, the recoverable information region can be reconfigured without phased-array beamforming, multiple RF chains, or mechanical motion. A prototype operating at 5.05~GHz is fabricated on a single-layer Rogers RO4350B substrate with an electrical footprint of $0.57 \times 0.47\lambda_{0}^{2}$. Measured 16-QAM communication results verify the reconfigurable information-beam behavior: in the selected E-plane cut, low bit error rate (BER) is confined to the intended information-beam sectors while off-beam angles exhibit large magnitude and phase errors, elevated BER, or unrecoverable constellations, despite high received SNR. In the measured H-plane cuts, low BER is maintained over nearly the full angular range, confirming that the antenna preserves omnidirectional information recovery in the orthogonal plane. These results demonstrate a compact planar route to reconfigurable directional modulation for secure wireless communication.

\end{abstract}


\section{Introduction}
\IEEEPARstart{E}{merging} wireless systems, including the Internet of Things (IoT), short-range links, low-complexity edge devices, and unmanned communication platforms, continue to increase the demand for security mechanisms that can be implemented with limited hardware overhead~\cite{5751298}. In many of these systems, conventional cryptographic solutions remain essential, but the required computation, latency, and protocol management can become undesirable for compact radio-frequency (RF) front ends and highly integrated wireless nodes. This has motivated strong interest in physical-layer security (PLS), where the propagation channel and antenna radiation itself are used to improve confidentiality~\cite{7467419,8509094}. Within this context, directional modulation (DM) has become a particularly attractive approach because it uses antenna radiation to generate direction-dependent constellation distortion, enabling correct demodulation only in intended angular regions while degrading reception elsewhere~\cite{5159486,5422702,6645431,6746064}. For compact wireless platforms, this direct connection between radiation behavior and communication security is especially appealing.

Most DM implementations reported to date, however, are based on phased arrays or pattern-reconfigurable arrays with multiple RF chains, phase shifters, and relatively complex feed networks~\cite{5159486,5422702,5439878,6544472,9184842}. These architectures provide strong spatial waveform control, but they also introduce increased cost, power consumption, calibration burden, and structural complexity. Such requirements are often incompatible with low-cost, compact, and planar wireless hardware. To simplify the implementation, several alternatives have been explored, including near-field direct antenna modulation~\cite{4684619}, distributed antenna dynamics~\cite{9665259,10217146}, vector or space--time modulation using 4-D antenna arrays~\cite{9664476}, and dynamic phase center control based on spatial amplitude modulation~\cite{10286341}. These works show that secure transmission can be achieved without relying on a conventional phased-array architecture, but most of them still require specialized platforms, array-level complexity, or nontrivial excitation networks.

More recently, simplified antenna-level realizations have begun to emerge. A single-element dynamic antenna for secure wireless applications was reported in~\cite{10161710}, demonstrating that dynamic waveform control can be implemented directly on a compact radiator. Dynamic directional modulation has also been studied using electrically small antennas for omnidirectional operation~\cite{9674846}, a compact stacked patch structure with $360^{\circ}$ beam steering~\cite{9140321}, and a dipole antenna with a dynamic balun for amplitude-based directional modulation~\cite{11030223}. These studies are highly relevant to the present work because they move DM toward simpler and more antenna-centric implementations. At the same time, they also indicate that an important challenge remains unresolved: it is still difficult to realize a compact, planar, and practically reconfigurable antenna that preserves omnidirectional coverage while producing a narrow information-recoverable region using a simple excitation strategy.

This challenge is central to the type of wireless platform considered in this paper. For short-range and mobile systems, omnidirectional coverage is often desirable to reduce alignment constraints, but secure transmission requires the information-bearing field to remain spatially selective. In addition, the antenna should be compact, low cost, planar, and compatible with simple RF front-end integration. These requirements motivate a dynamic antenna architecture that does not rely on conventional beam steering with multiple RF chains, but instead achieves information control directly through switching-induced aperture reconfiguration.

\begin{figure*}[!t]
	\begin{center}
		\includegraphics[width=7in]{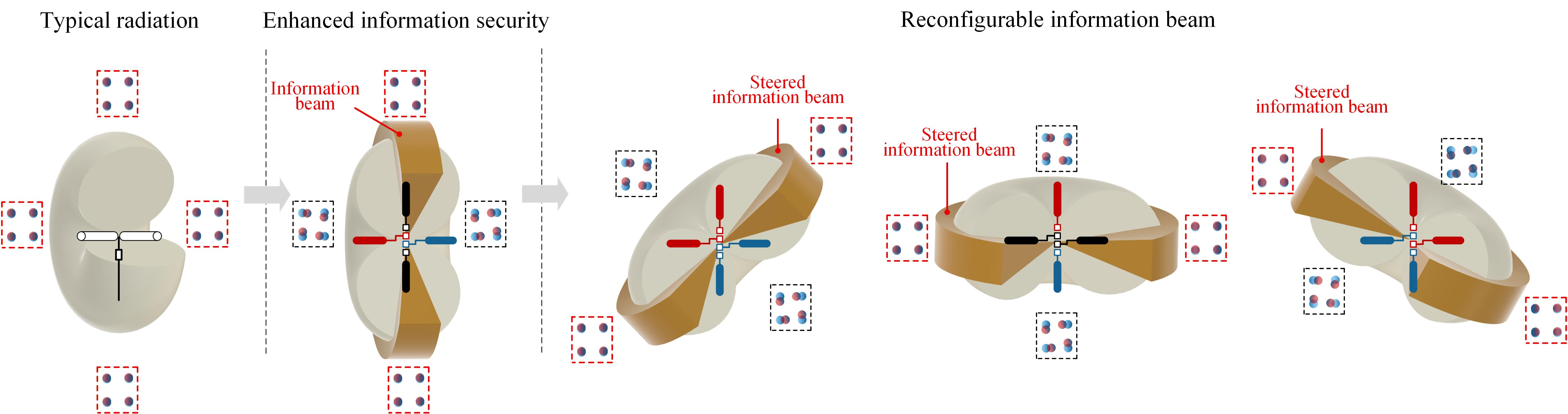}%
		
	\end{center}
	\caption{Conceptual topology of the proposed dynamic aperture-level directional modulation scheme. For small antennas, typical radiation distributes information uniformly over space, resulting in observable and recoverable signal constellations in all directions (left). By introducing dynamic switching at the antenna aperture, controlled differential currents are generated between antenna ports, which reconfigure the aperture phase pattern and give rise to a differential information-bearing radiation component (center). This mechanism forms a spatially confined information beam in which the modulation constellation remains recoverable at the intended receiver, while constellation distortion occurs outside the secure angular region. By reconfiguring the switching states, the information beam can be dynamically steered to different directions without modifying the physical antenna geometry or increasing the number of RF chains (right).}\label{overview}
\end{figure*}

In this paper, we present a compact reconfigurable dynamic antenna for secure wireless communication. The proposed structure uses four printed meander-line monopoles arranged in a planar topology and excited through a dynamic switching network. The key novelty is that the recoverable-information region is rotated by switching the odd-symmetric differential aperture component of a compact planar cross structure, which introduces apparent two-dimensional phase center displacement while the average radiation component remains broadly omnidirectional. Meander-line antennas are widely used for miniaturization because the folded current path increases electrical length while maintaining a compact footprint~\cite{5960759,6353124,7160701,5456169,10414395,5559341,99054,6171817}. Their compatibility with printed fabrication has led to many practical implementations in portable devices, wearable systems, RFID tags, and system-in-package antennas~\cite{1504825,9263312,9128053,6335459,5299014,6648416,9591355,7748548,7384427,5979187,6843861,8758307}. In the present work, these properties are exploited to realize a small and fully planar antenna platform that is suitable for dynamic information modulation.

The operating principle of the proposed antenna is illustrated in Fig.~\ref{overview}. Instead of using conventional beamforming, the proposed approach introduces dynamic switching at the antenna aperture so that the radiated field can be interpreted as the combination of an average radiation component and a differential information-bearing component. The switching operation creates controlled differential currents between antenna ports, reconfiguring the aperture phase pattern and producing angle-dependent changes in the transmitted constellation. As a result, the modulation remains recoverable only within a confined information beam, while receivers outside that region observe distorted constellation behavior and elevated bit error rate (BER). Importantly, this mechanism is achieved without changing the physical antenna geometry or adding multiple RF chains.

Based on this concept, a four-element broadside-oriented array is developed with an overall size of only $0.55 \times 1.73 \times 0.0045\,\lambda_{0}^{3}$ at 5~GHz. Dynamic radiation modulation is realized by switching the excitation paths and power ratios among the elements, which produces complementary E-plane phase responses together with differential magnitude patterns. This behavior leads to strongly angle-dependent BER in the selected E-plane cut, whereas the corresponding H-plane cuts remain quasi-static and broadly recoverable over the measured angular range. A digital four-path RF switching system implemented with commercial switches, phase shifters, attenuators, and power splitters provides a practical low-complexity realization.

The main contributions of this work are as follows. First, a compact planar reconfigurable dynamic antenna is developed for antenna-level DM using a single-layer meander-line topology and a simple switching network. Second, an average--differential array factor formulation is established to explain how controlled differential currents reconfigure the aperture phase pattern, produce apparent two-dimensional phase center displacement, and steer the information beam in four directions without phased-array beamforming. Third, the concept is experimentally validated by radiation and communication measurements, showing that reliable 16-QAM reception is confined to direction-dependent low-BER sectors in the selected E-plane cut, while omnidirectional low-BER recovery is maintained over nearly the full measured H-plane angular range. These results provide a practical route toward compact dynamic antennas for secure omnidirectional wireless communication.

The rest of this paper is organized as follows. Section~II presents the design of the compact planar meander-line antenna and the dynamic four-element array, together with the theoretical analysis of the switching-induced radiation behavior. Section~III investigates the communication performance, including angular BER characteristics and experimental validation. Finally, Section~IV concludes the paper.

\section{The Design of Reconfigurable Dynamic Antenna}
\subsection{Single Meander Line Monopole}
\begin{figure*}[!t]
	\centering
	\subfloat[]{\includegraphics[width=2.2in]{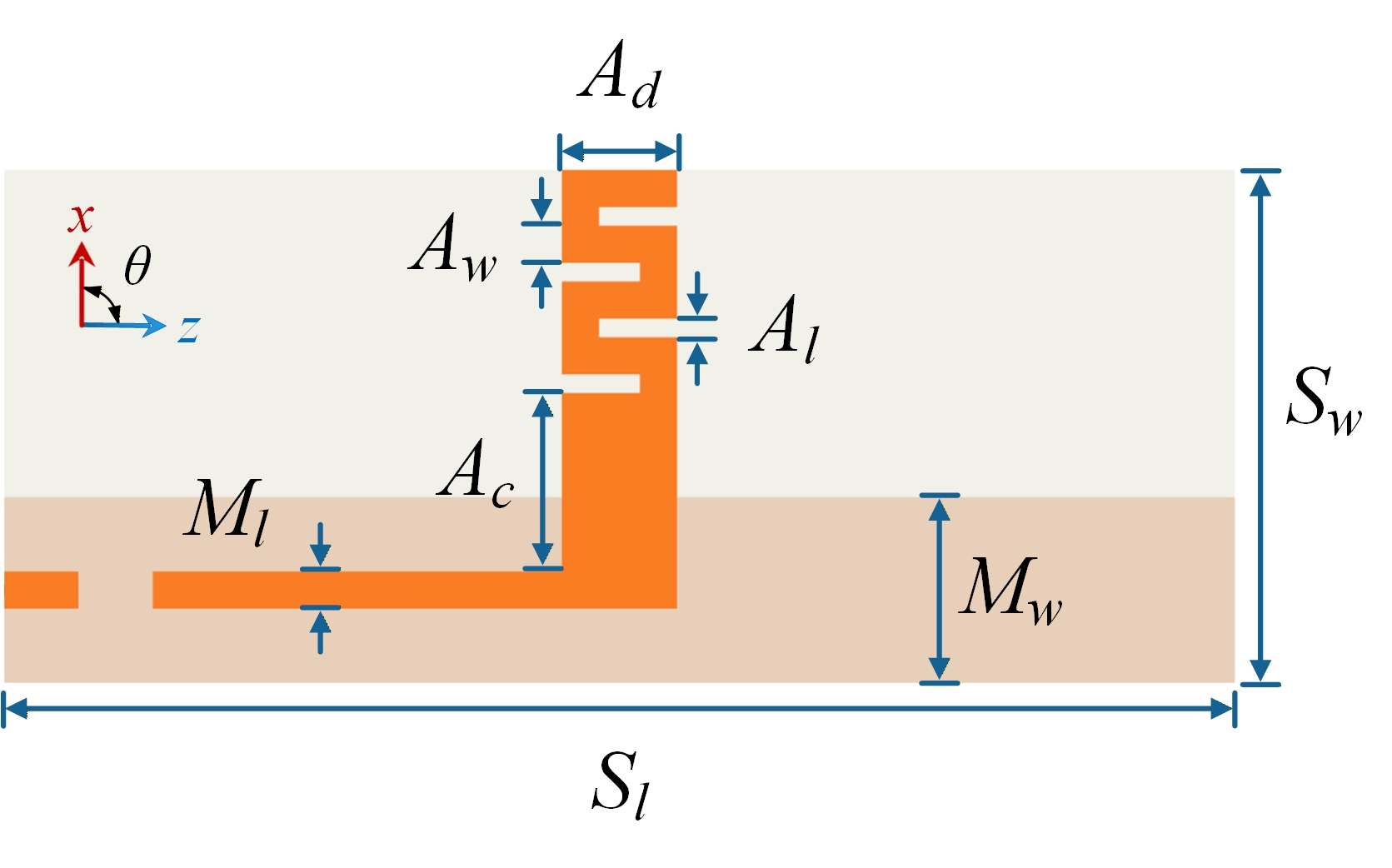}%
	}\hfil
	\subfloat[]{\includegraphics[width=2.4in]{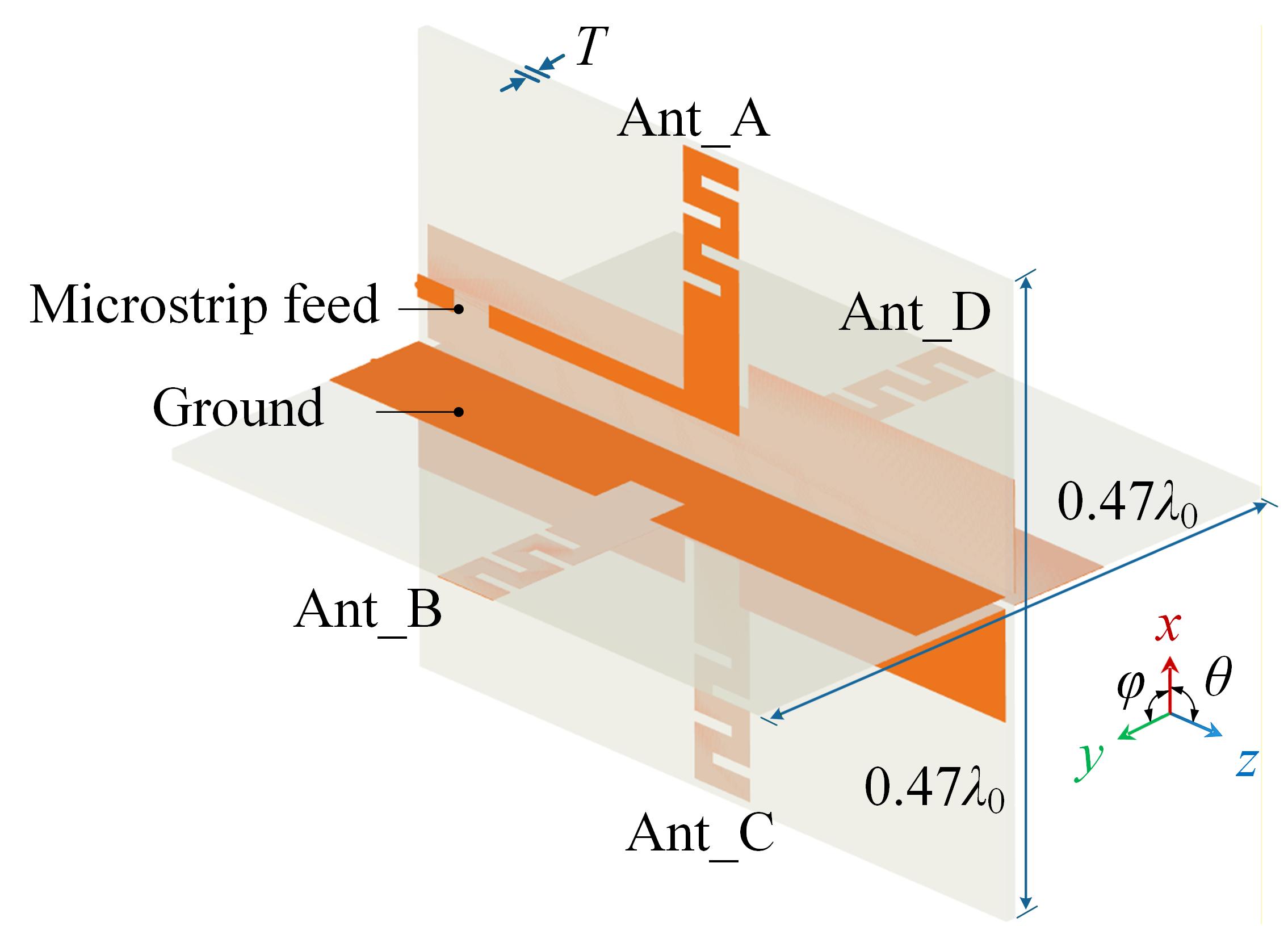}%
	}\hfil
	\subfloat[]{\includegraphics[width=1.8in]{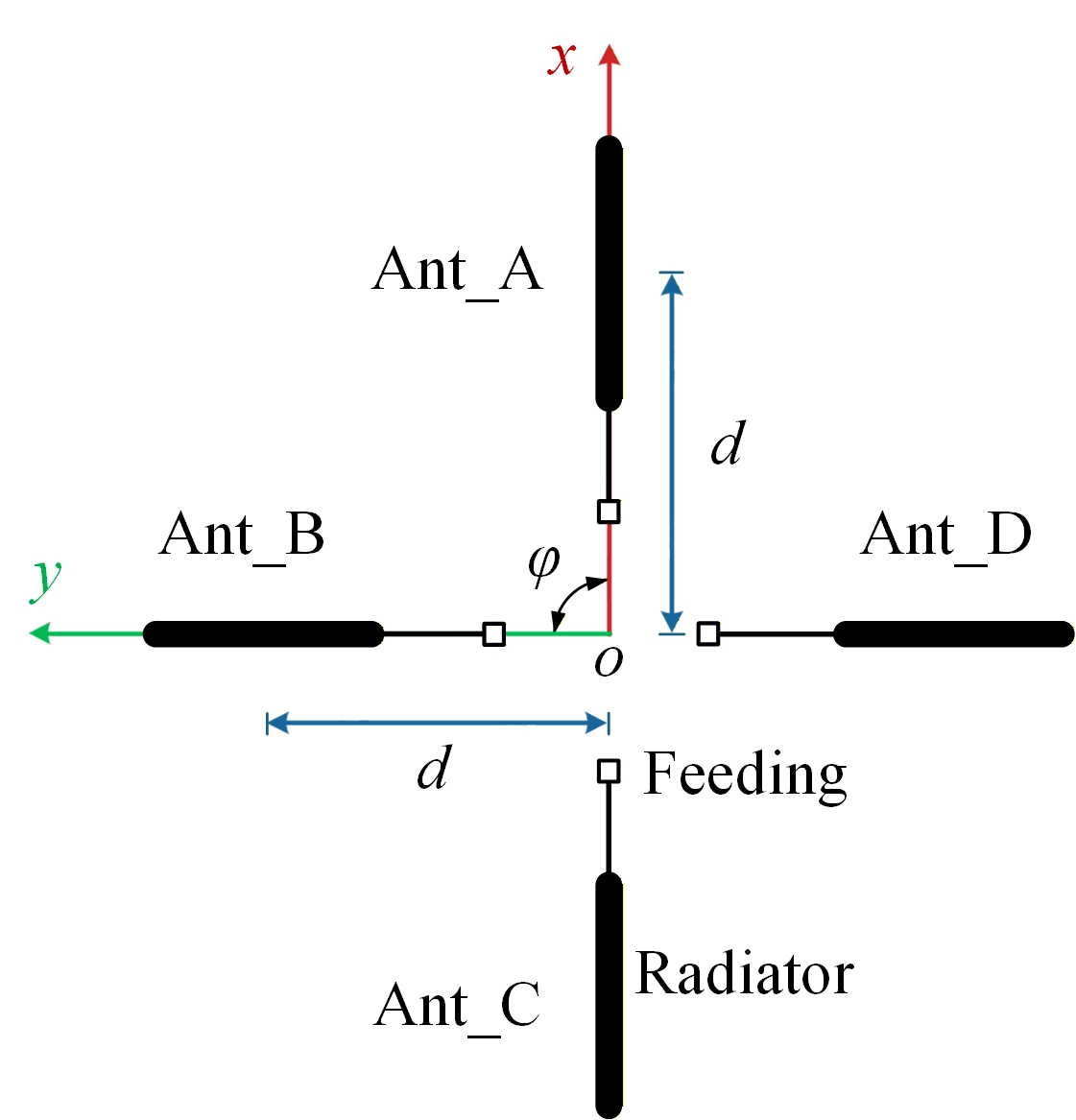}%
	}\caption{Geometry and topology of the proposed compact monopole-based antenna system.
		(a) Geometry and key dimensions of the meandered monopole radiator.
		(b) Three-dimensional view of the four-element antenna topology for reconfigurable information beam, illustrating the compact and symmetric configuration.
		(c) Equivalent antenna model used for theoretical analysis, where four monopole radiators (Ant~A--D) are arranged orthogonally around origin with spacing $d$, and the azimuth angle $\varphi$ defines the observation direction.}\label{fig:antenna_geometry}
\end{figure*}

\begin{table}[t]
	\caption{Optimized Dimensions of the Planar Meander-Line Antenna for 5~GHz Operation (Unit: mm)}
	\label{tab:table1}
	\centering
	\scriptsize
	\setlength{\tabcolsep}{3pt}
	\renewcommand{\arraystretch}{1.1}
	\begin{tabular}{cccccccccc}
		\hline\hline
		$A_l$ & $A_w$ & $A_d$ & $M_l$ & $M_w$ & $S_l$ & $T$ & $A_c$ & $S_w$ & Gain (dBi) \\
		\hline
		0.55 & 1 & 4 & 1.08 & 5 & 34 & 0.508 & 5 & 14.2 & 2.10 \\
		\hline\hline
	\end{tabular}
\end{table}
The design methodology of the meander-line radiator is well developed in
\cite{99054,10379428}, where the meander structure is modeled as two
equivalent parts: a folded short-circuited terminal circuit and an
equivalent straight conductor. Self-resonance is achieved through the
balance between the capacitive reactance introduced by the short dipole
section and the inductive reactance associated with the folded segments.
The design of the proposed dynamic array begins with a planar meander-line
monopole element, as shown in Fig.~\ref{fig:antenna_geometry}(a). To achieve low-cost fabrication
and a compact implementation, the entire antenna structure---including the
meander-line radiating element, capacitive patch, microstrip feed, and
ground plane---is realized on a single layer of Rogers RO4350B substrate
with a relative permittivity of 3.48 and a dissipation factor of 0.0037.
A standard substrate thickness of $T=0.508$~mm is selected to ensure a
compact and lightweight structure. The antenna is excited by a 50-$\Omega$
microstrip line with width $M_{l}$. A capacitive patch with length $A_{c}$,
inspired by the design reported in \cite{4020418}, is incorporated to
further miniaturize the antenna and to provide a practical impedance
matching solution. The proposed meander-line monopole employs two meander sections, with
the line width and gap defined by $A_{w}$ and $A_{l}$, respectively. The
capacitive patch and the meander-line radiator share the same overall
width $A_{d}$. To preserve omnidirectional radiation characteristics, the
ground plane dimensions, denoted by $S_{l}$ and $M_{w}$, are optimized such
that the maximum realized gain is orthogonal to the electrical
polarization. All geometrical parameters are optimized for resonance at
5~GHz and are summarized in Table~\ref{tab:table1}.

Four identical meander-line monopole elements are subsequently arranged
orthogonally and excited through a shared feeding structure, forming a
compact and symmetric topology as illustrated in Fig.~\ref{fig:antenna_geometry}(b), with the equivalent model shown in Fig.~\ref{fig:antenna_geometry}(c). This
configuration preserves omnidirectional radiation characteristics while
providing an effective physical platform for the differential-current-
based directional modulation scheme analyzed in this work.

\begin{figure}[t]
	\centering
	\includegraphics[width=3in]{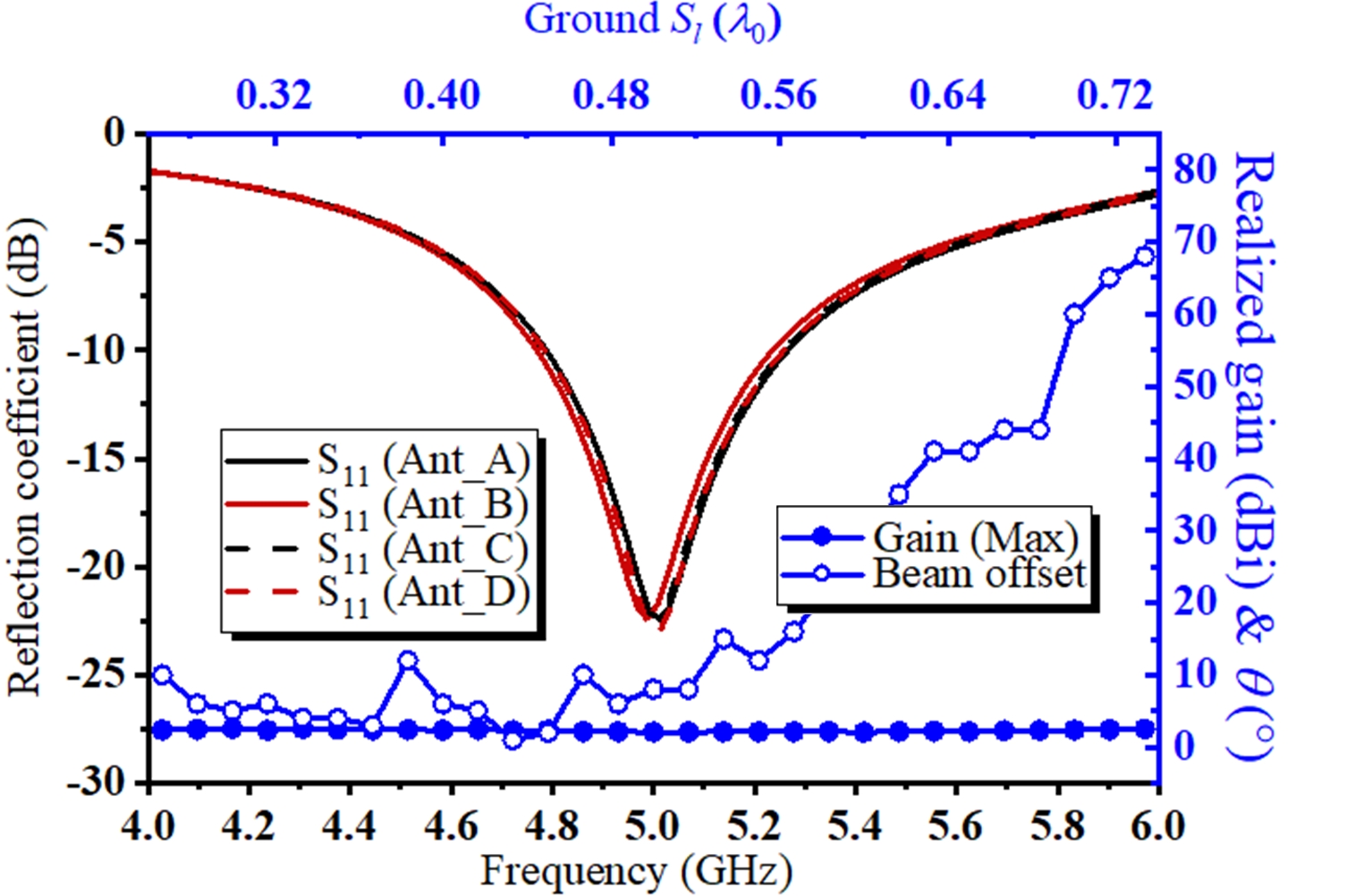}%
	\caption{The simulated ground plane effects on radiation characteristics.}\label{fig:ground_plane}
\end{figure}

The simulated ground-plane effect is summarized in Fig.~\ref{fig:ground_plane}, where the reflection coefficient, realized gain, radiation efficiency, and main-beam direction are evaluated as the ground length is varied. This study is included because the proposed antenna should preserve an omnidirectional azimuth-plane response while maintaining its strongest radiation near broadside.
In Fig.~\ref{fig:ground_plane}, the maximum realized gain and beam offset are analyzed for different ground lengths \(S_{l}\), while the ground width \(M_{w}\) is maintained at 5~mm to support the microstrip transition. The maximum realized gain remains near 2.3~dBi when good impedance matching is maintained. The beam-offset trend shows that the ground length should remain below approximately one half wavelength to preserve broadside radiation.

\subsection{Two-State Switching Array Factor Formulation and Four-Direction IB Steering}

Using the standard far-field factorization from array theory~\cite{balanis2016antenna}, the radiated field is expressed as
\begin{equation}
	E(\theta,\varphi,t)=C\,a(\theta,\varphi)\,AF(\theta,\varphi,t),
\end{equation}
where $C$ contains the free-space propagation terms, $a(\theta,\varphi)$ is the element factor, and $AF(\theta,\varphi,t)$ is the time-varying array factor.

Consider a four-element array placed in the $x$--$y$ plane with the array center at the origin (Fig.~\ref{fig:antenna_geometry}(c)). We adopt the azimuth convention that $\varphi=0^\circ$ points along the $+x$ axis. Let $d$ denote the effective phase center distance from the array center to each element:
\begin{subequations}
	\begin{align}
		\mathbf{A} &= (d,0,0), \label{eq:A}\\
		\mathbf{B} &= (0,d,0), \label{eq:B}\\
		\mathbf{C} &= (-d,0,0), \label{eq:C}\\
		\mathbf{D} &= (0,-d,0). \label{eq:D}
	\end{align}
\end{subequations}
Using spherical coordinates, the far-field observation direction is
\begin{equation}
	\hat{\mathbf r}=
	\big(\sin\theta\cos\varphi,\ \sin\theta\sin\varphi,\ \cos\theta\big).
\end{equation}
\begin{table*}[t]
	\centering
	\caption{Summary of dynamic switching modes, switching states, and resulting array factors}
	\label{tab:table2}
	\renewcommand{\arraystretch}{1.25}
	\begin{tabular}{c c c c c c}
		\hline\hline
		Dynamic mode
		& $\mathbf{s}^{(1)}$
		& $\mathbf{s}^{(2)}$
		& $AF_{\mathrm{avg}}$
		& $AF_{\Delta}$
		& Realized gain (dBi) \\
		\hline
		$\varphi=0^\circ$
		& $[1,0,0,0]$
		& $[0,0,1,0]$
		& $\cos\psi_x$
		& $j\sin\psi_x$
		& 1.93 \\
		
		$\varphi=45^\circ$
		& $[1,1,0,0]$
		& $[0,0,1,1]$
		& $\cos\psi_x+\cos\psi_y$
		& $j(\sin\psi_x+\sin\psi_y)$
		& 2.04 \\
		
		$\varphi=90^\circ$
		& $[0,1,0,0]$
		& $[0,0,0,1]$
		& $\cos\psi_y$
		& $j\sin\psi_y$
		& 1.95 \\
		
		$\varphi=135^\circ$
		& $[0,1,1,0]$
		& $[1,0,0,1]$
		& $\cos\psi_x+\cos\psi_y$
		& $j(\sin\psi_y-\sin\psi_x)$
		& 2.11 \\
		\hline\hline
	\end{tabular}
	
	\vspace{2pt}
	\parbox{\linewidth}{\textit{Note:} In the switching state vectors, `1' and `0'
		denote strong and weak excitations of the corresponding antenna elements,
		respectively. The realized gain values correspond to both complementary
		switching states and indicate that dynamic switching preserves the
		omnidirectional power radiation.}
\end{table*}

\begin{figure*}[!t]
	\centering
	\subfloat[]{\includegraphics[width=3.2in]{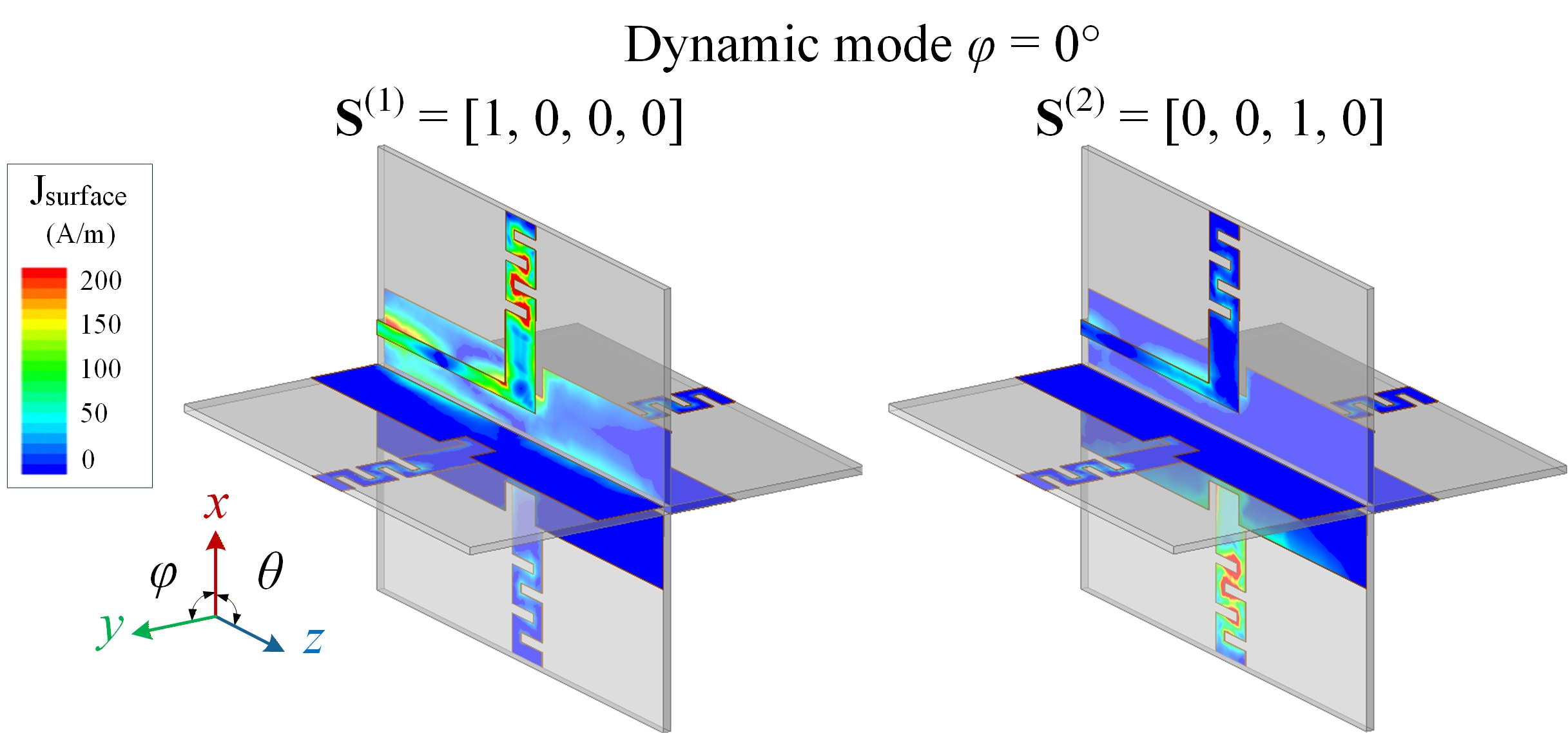}}\hfil
	\subfloat[]{\includegraphics[width=3.2in]{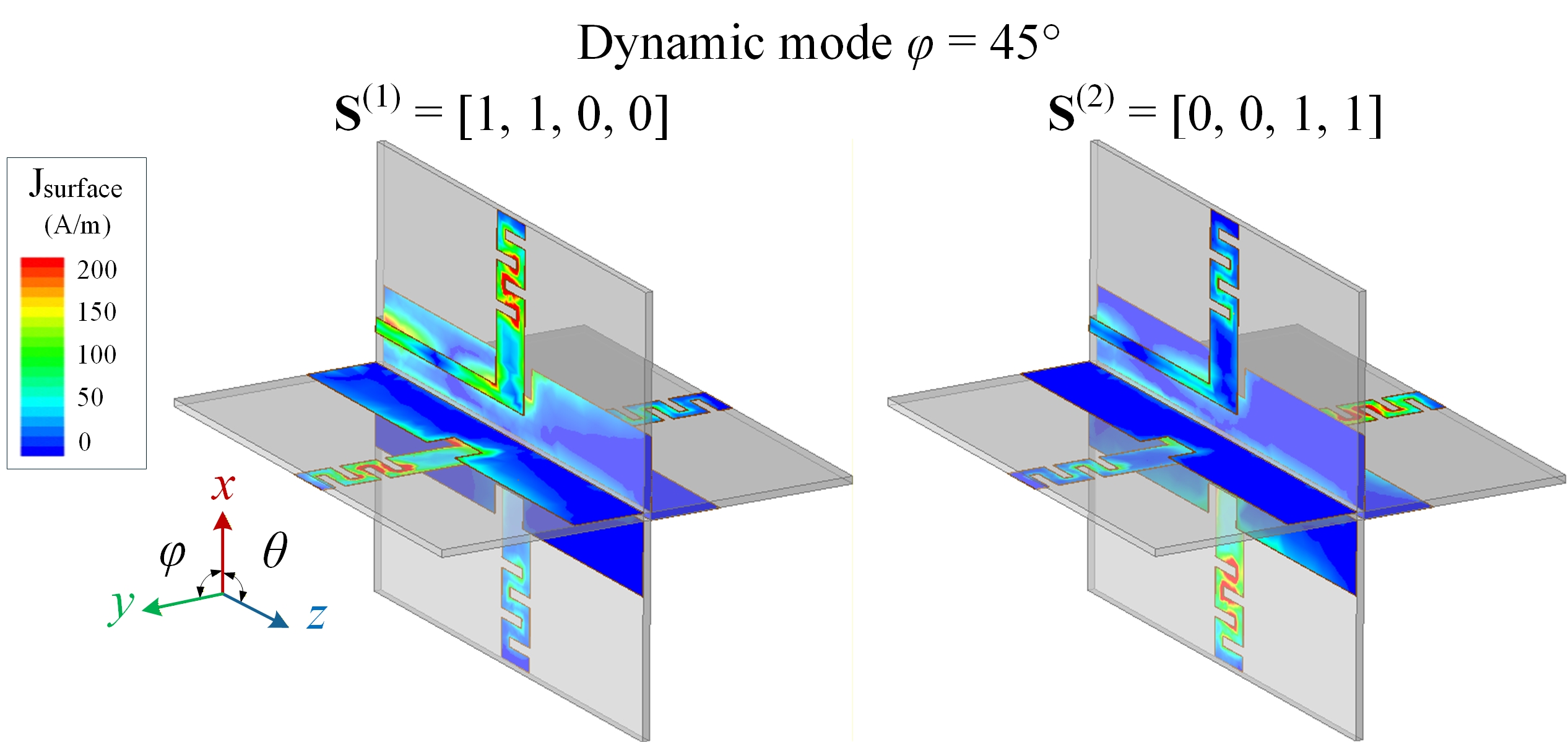}}\hfil\\
	\subfloat[]{\includegraphics[width=3.2in]{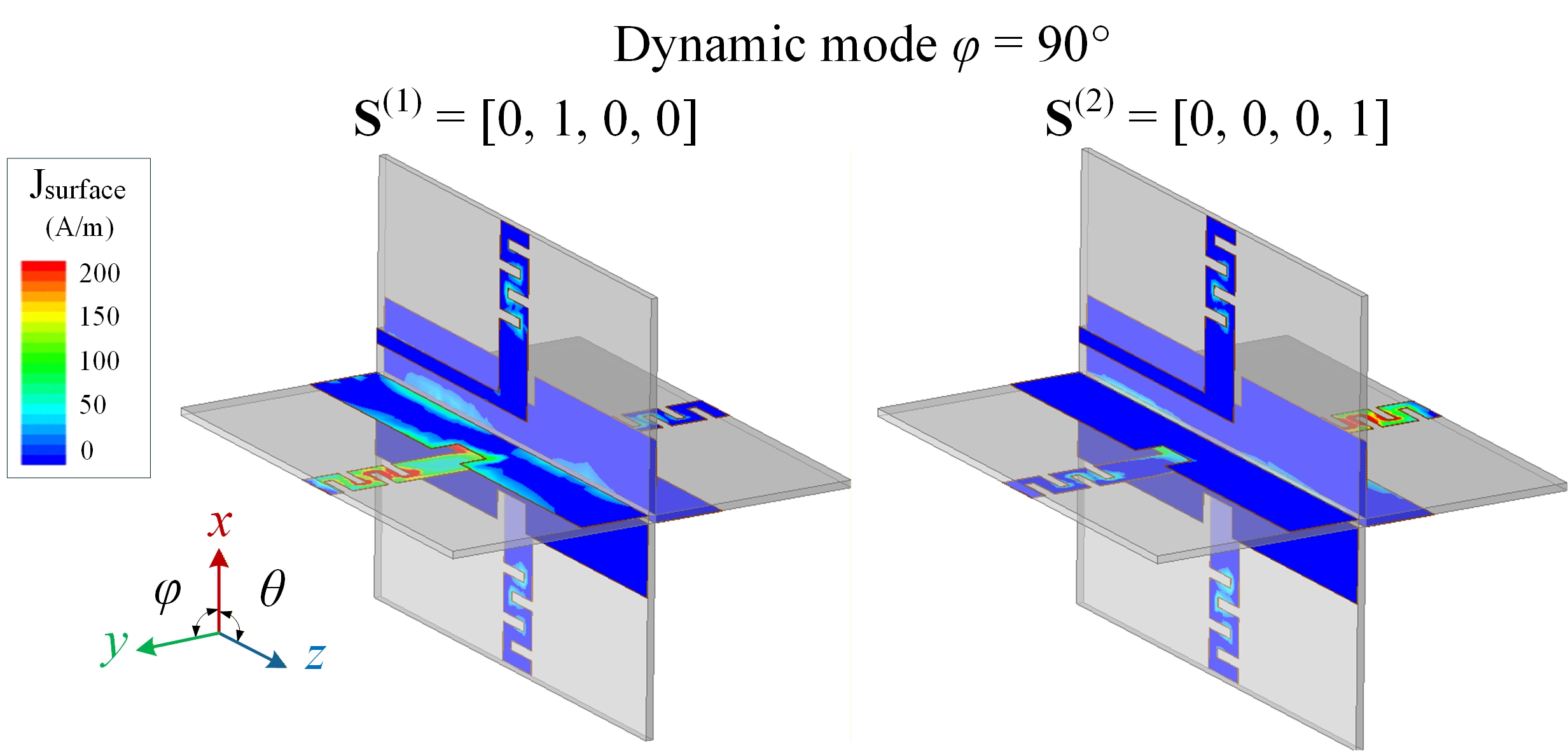}}\hfil
	\subfloat[]{\includegraphics[width=3.2in]{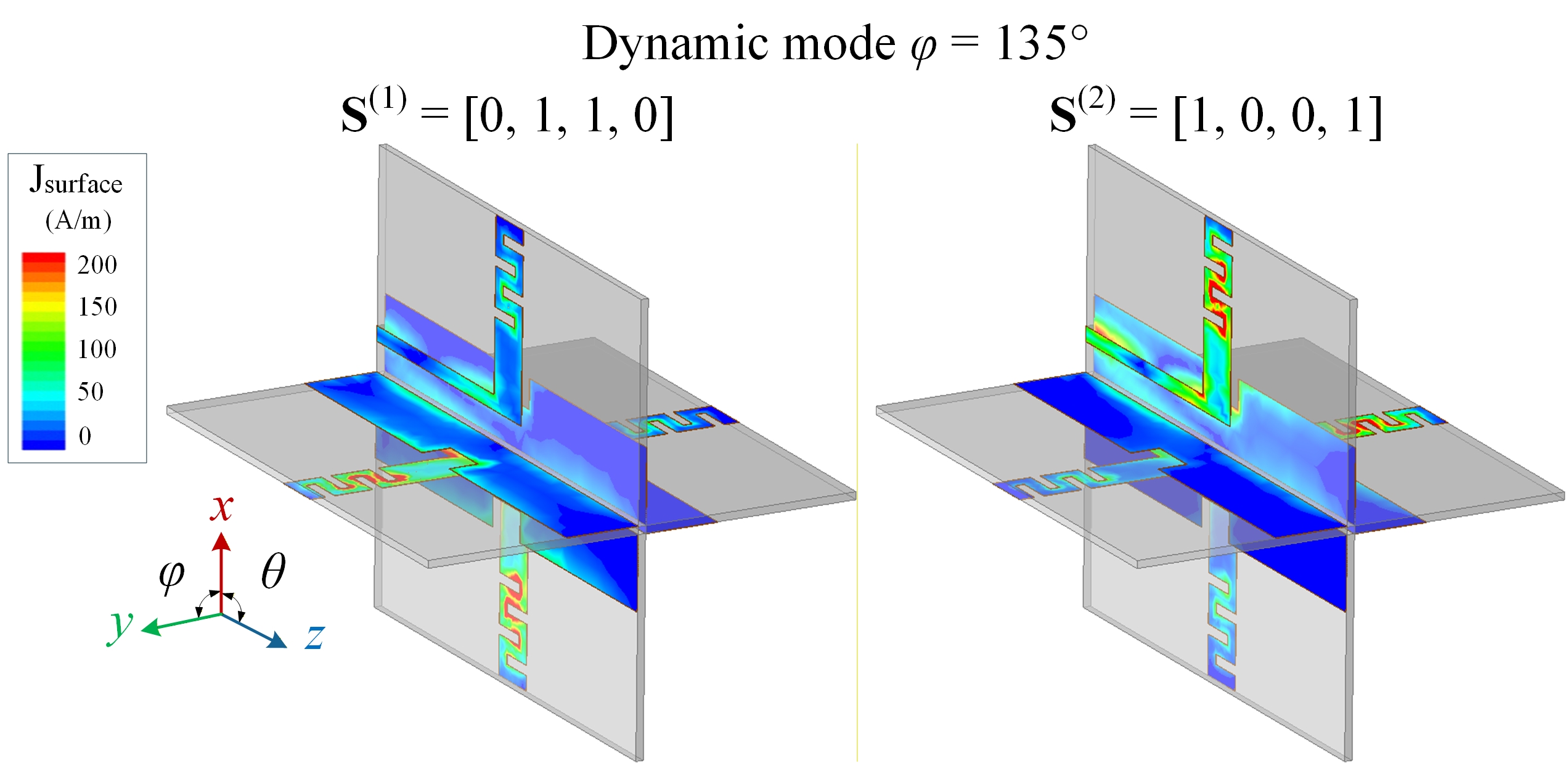}}\hfil
		\subfloat[]{\includegraphics[width=1.7in]{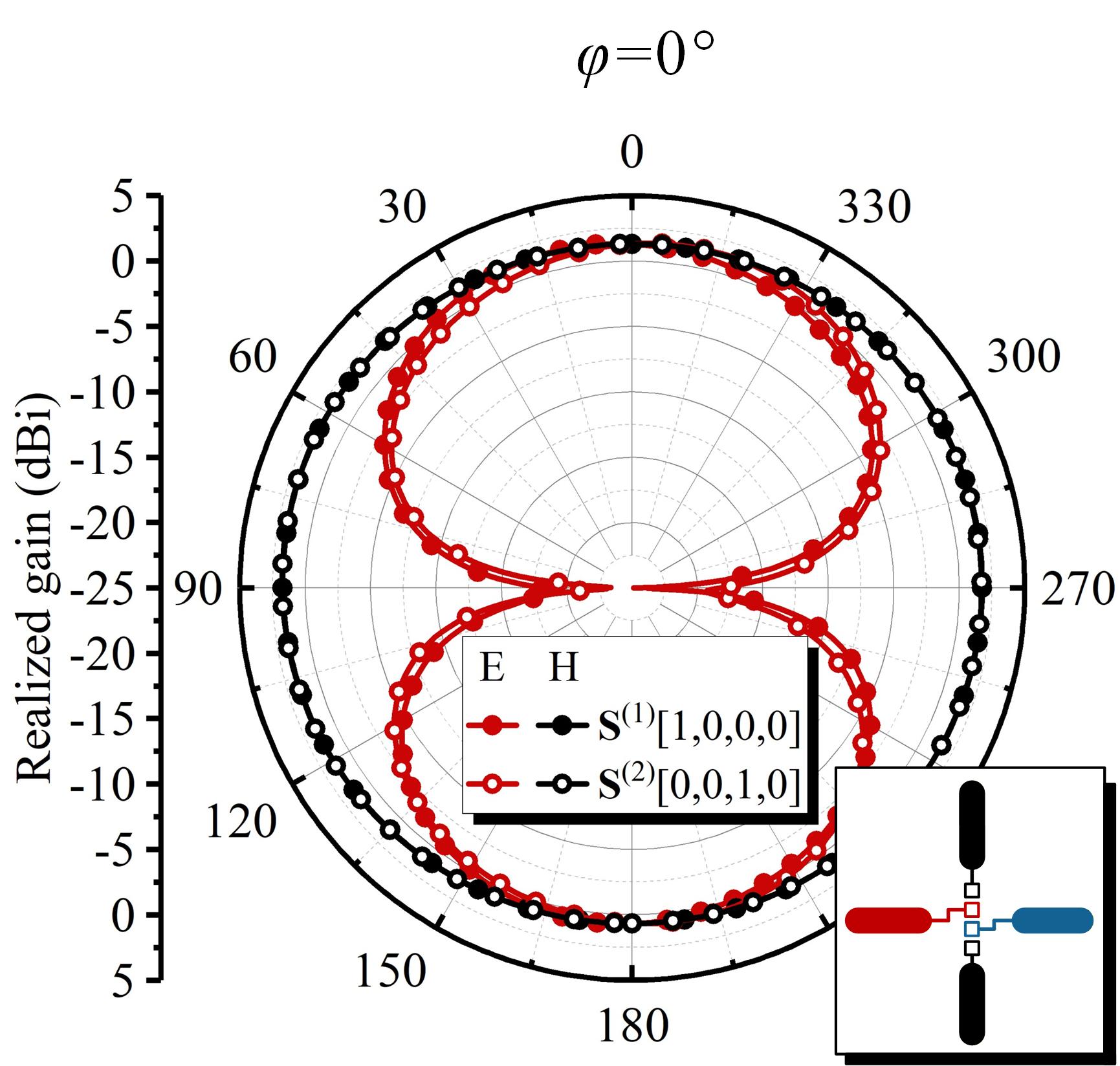}}\hfil
		\subfloat[]{\includegraphics[width=1.7in]{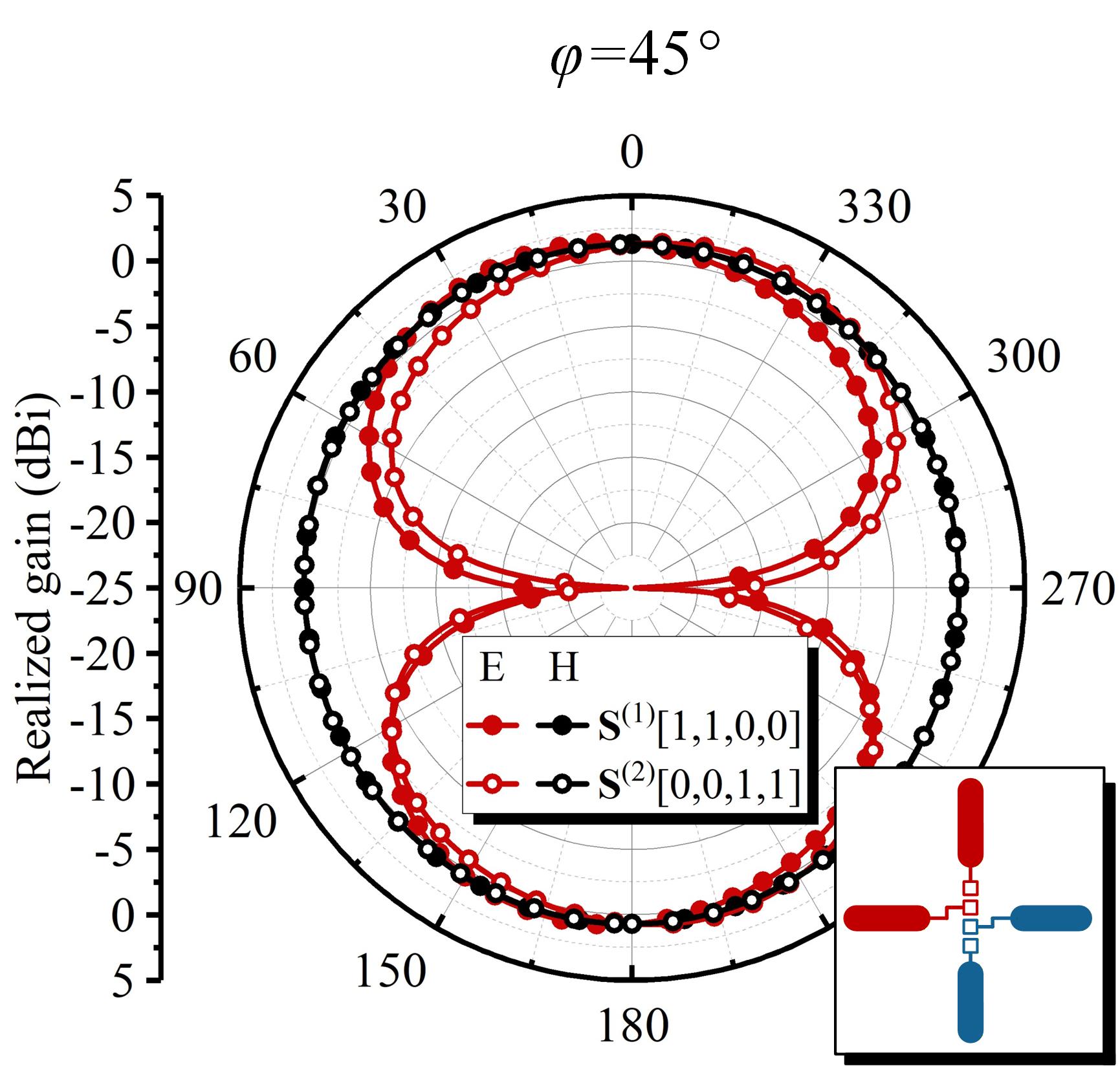}}\hfil
		\subfloat[]{\includegraphics[width=1.7in]{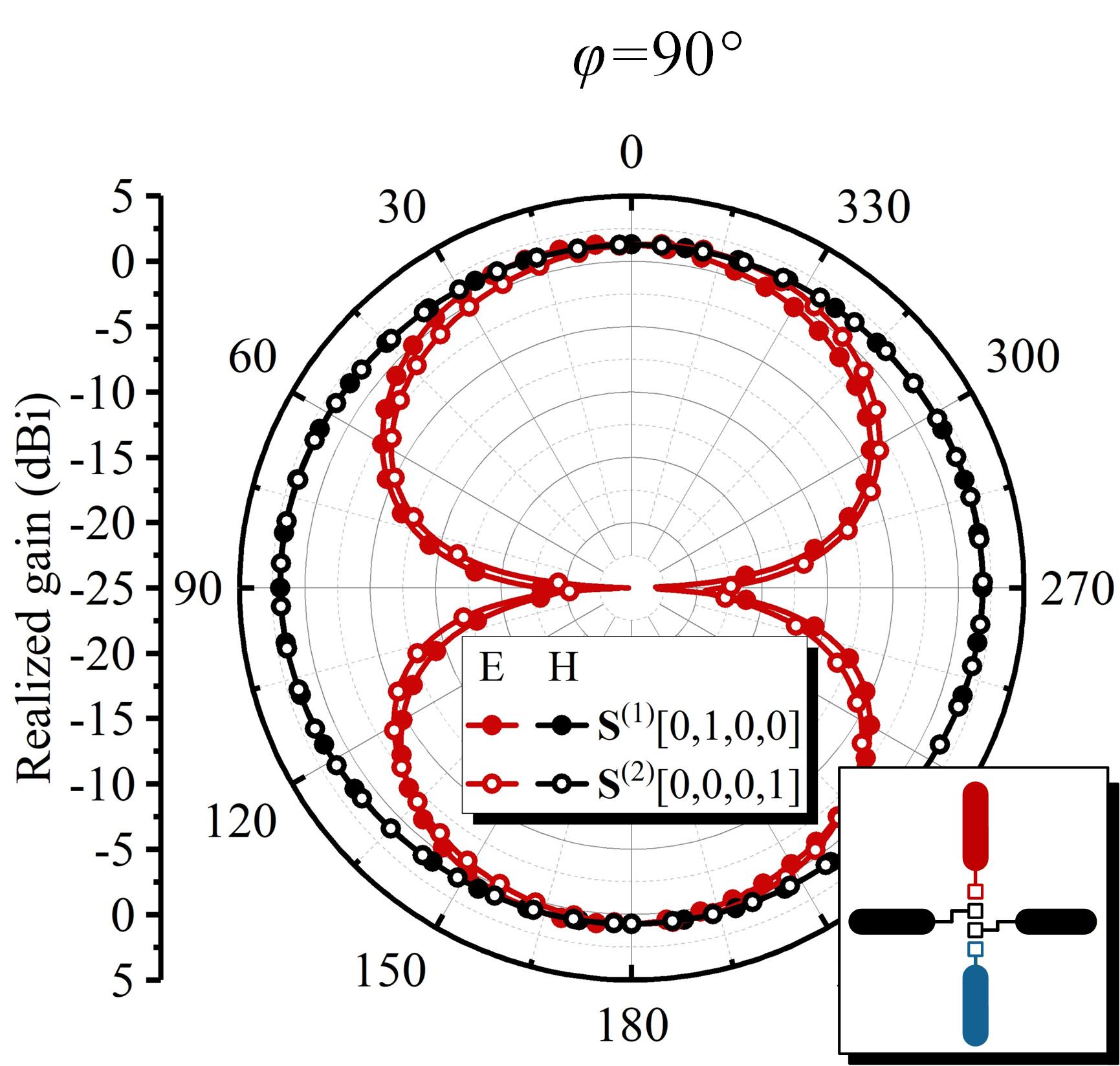}}\hfil
		\subfloat[]{\includegraphics[width=1.7in]{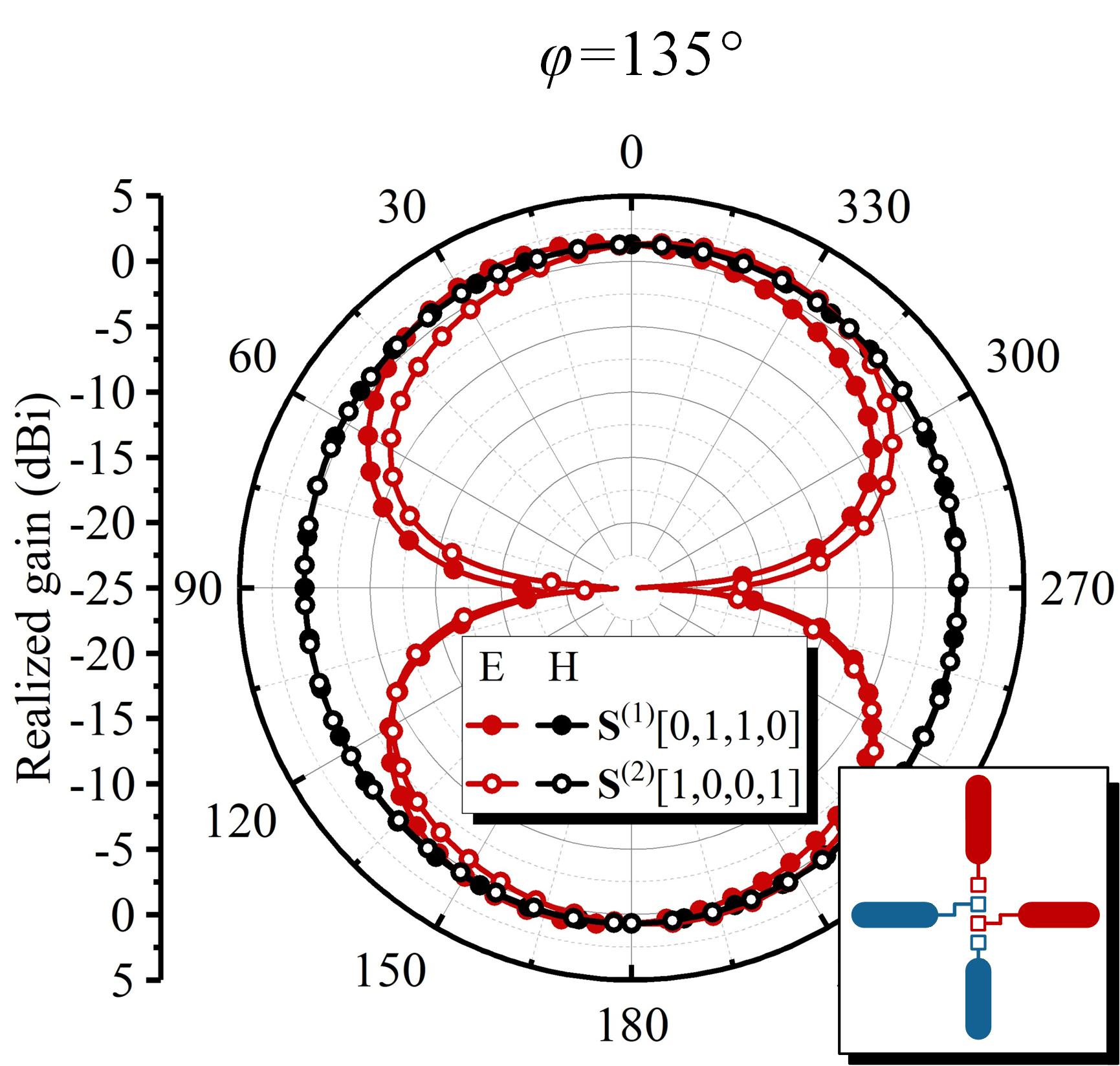}}\hfil
	\caption{Simulated surface current distributions and simulated realized gain patterns
		of the proposed antenna under different dynamic switching modes with a switching power ratio of $\alpha=9$~dB. (a)--(d) Surface current distributions of the two complementary switching states $\mathbf{s}^{(1)}$ and $\mathbf{s}^{(2)}$ for dynamic modes $\varphi=0^\circ$, $45^\circ$, $90^\circ$, and $135^\circ$, respectively. For each dynamic mode, the two states exhibit distinct odd-symmetric current distributions across the aperture, corresponding to differential port currents with a $180^\circ$ phase difference. (e)--(h) Corresponding realized gain patterns in the E-plane and H-plane, showing that the radiation power distributions of the two switching states remain nearly identical. These results confirm that the proposed directional modulation and reconfigurable information beam steering are achieved through phase-pattern reconfiguration rather than power redistribution.}
	\label{fig:dynamic_patterns}
\end{figure*}

The phase term associated with the $n$th element is $e^{jk\mathbf r_n\cdot\hat{\mathbf r}}$. For the chosen geometry, it is convenient to define two scalar phase variables:
\begin{subequations}
	\begin{align}
		\psi_x = k d \sin\theta\cos\varphi, \label{eq:psix}\\
		\psi_y = k d \sin\theta\sin\varphi, \label{eq:psiy}
	\end{align}
\end{subequations}
so that the four elements contribute phase factors
\begin{subequations}
	\begin{align}
		e^{j k \mathbf{r}_A \cdot \hat{\mathbf{r}}} &= e^{+j\psi_x}, \label{eq:phaseA}\\
		e^{j k \mathbf{r}_B \cdot \hat{\mathbf{r}}} &= e^{+j\psi_y}, \label{eq:phaseB}\\
		e^{j k \mathbf{r}_C \cdot \hat{\mathbf{r}}} &= e^{-j\psi_x}, \label{eq:phaseC}\\
		e^{j k \mathbf{r}_D \cdot \hat{\mathbf{r}}} &= e^{-j\psi_y}. \label{eq:phaseD}
	\end{align}
\end{subequations}
Let $w_n(t)=I_n(t)/I_0$ denote the normalized current weight of element $n\in\{A,B,C,D\}$ relative to a reference current $I_0$. The array factor is then written as the weighted sum of spatial phasors:
\begin{equation}
	\begin{aligned}
		AF(\theta,\varphi,t)
		&=
		w_A(t)e^{+j\psi_x}
		+ w_B(t)e^{+j\psi_y} \notag\\
		&\quad
		+ w_C(t)e^{-j\psi_x}
		+ w_D(t)e^{-j\psi_y}.
	\end{aligned}
\end{equation}
We consider two instantaneous switching states, denoted as ``state 1'' and ``state 2''. Each state is represented by a binary excitation vector
\begin{subequations}
	\begin{align}
		\mathbf{s}^{(1)} &=
		\big[s_A^{(1)},\, s_B^{(1)},\, s_C^{(1)},\, s_D^{(1)}\big], \label{eq:s1}\\
		\mathbf{s}^{(2)} &=
		\big[s_A^{(2)},\, s_B^{(2)},\, s_C^{(2)},\, s_D^{(2)}\big] \label{eq:s2}
	\end{align}
\end{subequations}
with
\begin{equation}
	s_n^{(i)}\in\{0,1\},\qquad \ i\in\{1,2\}.
\end{equation}
Here, $s_n^{(i)}=1$ indicates element $n$ is active (dominant) in state $i$, while $s_n^{(i)}=0$ indicates it is inactive (not dominant). To express two-state switching with a single waveform, we can define a square-wave variable
\begin{equation}
	\sigma(t)\in\{+1,-1\},
\end{equation}
where $\sigma(t)=+1$ selects state 1 and $\sigma(t)=-1$ selects state 2. Then each element weight can be written in a unified form:
\begin{equation}
	w_n(t)=\frac{1+\sigma(t)}{2}s_n^{(1)}+\frac{1-\sigma(t)}{2}s_n^{(2)},
\end{equation}
The above expression is a compact way to represent a two-point (discrete) switching distribution, analogous to two-level amplitude switching. Let $AF_1$ and $AF_2$ denote the array factors obtained by substituting $\mathbf{s}^{(1)}$ and $\mathbf{s}^{(2)}$ into the weighted sum, respectively. The switched array factor is then
\begin{equation}
	\begin{aligned}
		AF(\theta,\varphi,t)
		&= \frac{AF_1 + AF_2}{2}
		+ \sigma(t)\frac{AF_1 - AF_2}{2} \\
		&= AF_{\mathrm{avg}}(\theta,\varphi)
		+ \sigma(t)\,AF_{\Delta}(\theta,\varphi).
	\end{aligned}
\end{equation}
In this form, $AF_{\mathrm{avg}}$ corresponds to the time-averaged spatial response, while $AF_{\Delta}$ quantifies the additional angle-dependent modulation introduced by switching.

Using Euler's identity, the opposite-pair contributions can be written directly in terms of even cosine terms and odd sine terms. For either instantaneous state $i\in\{1,2\}$,
\begin{equation}
	\begin{aligned}
		AF_i(\theta,\varphi)=\;&
		\big(s_A^{(i)}+s_C^{(i)}\big)\cos\psi_x
		+\big(s_B^{(i)}+s_D^{(i)}\big)\cos\psi_y \\
		&+j\Big[
		\big(s_A^{(i)}-s_C^{(i)}\big)\sin\psi_x
		+\big(s_B^{(i)}-s_D^{(i)}\big)\sin\psi_y
		\Big].
	\end{aligned}
\end{equation}
This representation clearly separates the even (cosine) and odd (sine) spatial terms contributed by each opposite pair.

In this work, we introduce four dynamic modes that correspond to four principal steering directions in azimuth: $\varphi=0^\circ$, $45^\circ$, $90^\circ$, and $135^\circ$. Each mode is specified by the pair of binary state vectors $(\mathbf s^{(1)},\mathbf s^{(2)})$ and yields closed-form expressions for $AF_{\mathrm{avg}}$ and $AF_{\Delta}$.
As a representative example, consider dynamic mode 1, in which opposite-pair switching occurs
along the $x$ axis ($\varphi=0^\circ$), where the excitation alternates
between elements $A$ and $C$. In this case,
\begin{subequations}
	\begin{align}
		\mathbf{s}^{(1)} &= [1,\,0,\,0,\,0], \label{eq:s_state1}\\
		\mathbf{s}^{(2)} &= [0,\,0,\,1,\,0]. \label{eq:s_state2}
	\end{align}
\end{subequations}
The instantaneous array factors are $AF_1=e^{j\psi_x}$ and $AF_2=e^{-j\psi_x}$, giving
\begin{subequations}
	\begin{align}
		AF_{\mathrm{avg}} &= \cos\psi_x,\\
		AF_{\Delta} &= j\sin\psi_x,
	\end{align}
\end{subequations}
leading to
\begin{equation}
	AF(t)=\cos\psi_x+\sigma(t)\,j\sin\psi_x.
\end{equation}
Other dynamic switching modes follow the same formulation and differ
only in the angular dependence of $AF_{\Delta}$, corresponding to
opposite-pair switching along the $y$ axis ($\varphi=90^\circ$) and
diagonal switching along $\varphi=\pm45^\circ$. All related terms of four switching modes are calculated in Table~\ref{tab:table2}. For multi-element switching states such as $\mathbf{s}=[1,1,0,0]$ and
$\mathbf{s}=[0,1,1,0]$, the differential array factor consists of the
superposition of two orthogonal odd-symmetric components associated with
the $x$- and $y$-directed monopole pairs. Specifically,
$AF_{\Delta}\propto \sin\psi_x+\sin\psi_y$ and
$AF_{\Delta}\propto \sin\psi_y-\sin\psi_x$, respectively. In the azimuth
plane, this superposition leads to constructive interference along
$\varphi=45^\circ$ and $\varphi=135^\circ$, resulting in a rotation of
the information beam toward the corresponding diagonal directions. This
behavior confirms that the information beam steering is achieved through
superposition of differential radiation patterns rather than by activating
a single antenna.

Fig.~\ref{fig:dynamic_patterns} presents the simulated surface current distributions for the two complementary switching states of each dynamic mode. In all four cases, the alternating excitation gives rise to pronounced odd-symmetric current distributions across the antenna aperture, where currents at spatially symmetric locations exhibit equal magnitude but opposite directions. This behavior reflects the presence of differential port currents with an inherent $180^\circ$ phase difference, which is a fundamental characteristic of the monopole-based radiator and is required to generate a pure differential radiation component.

\begin{figure*}[t]
	\centering
	\subfloat[]{\includegraphics[width=3.2in]{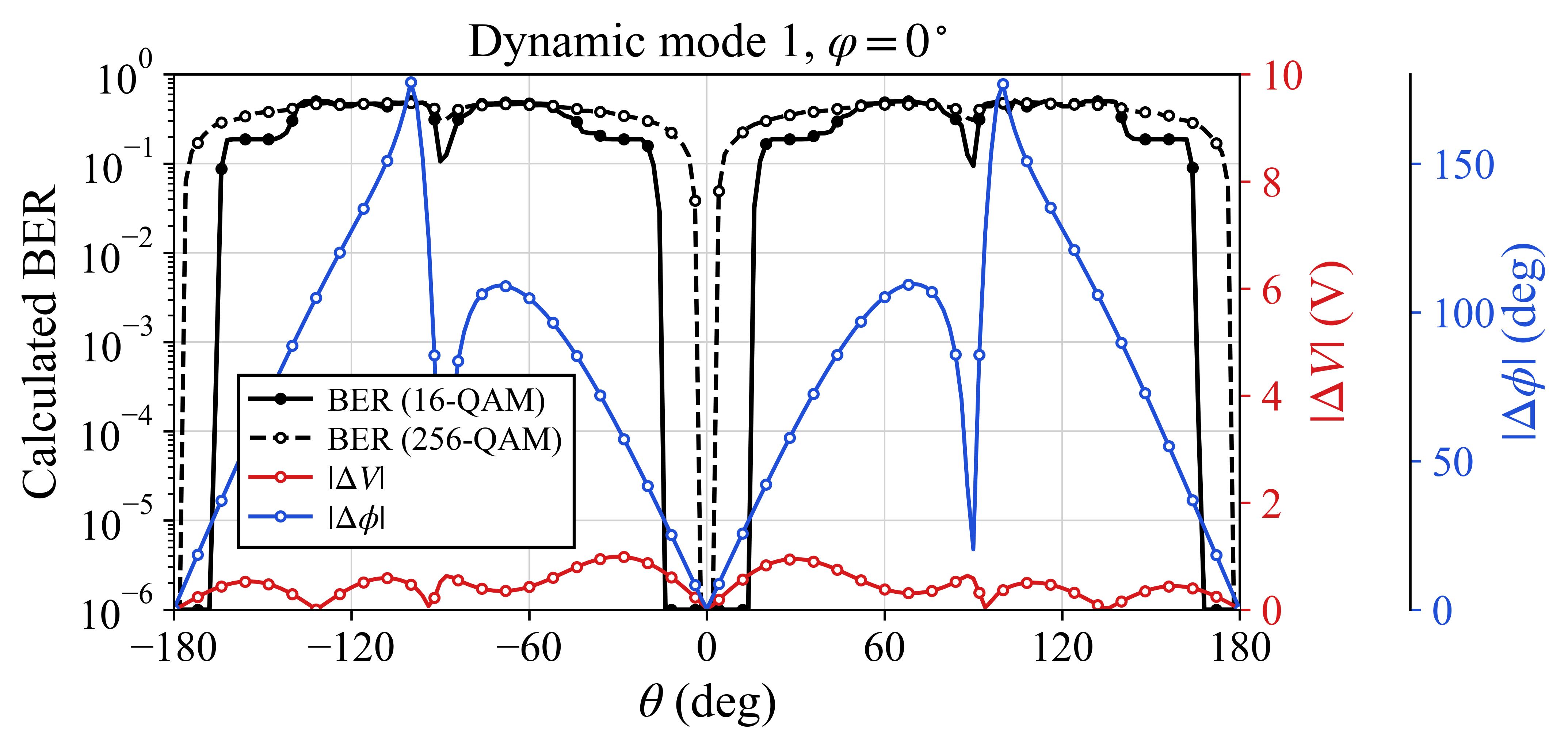}}\hfil
	\subfloat[]{\includegraphics[width=3.2in]{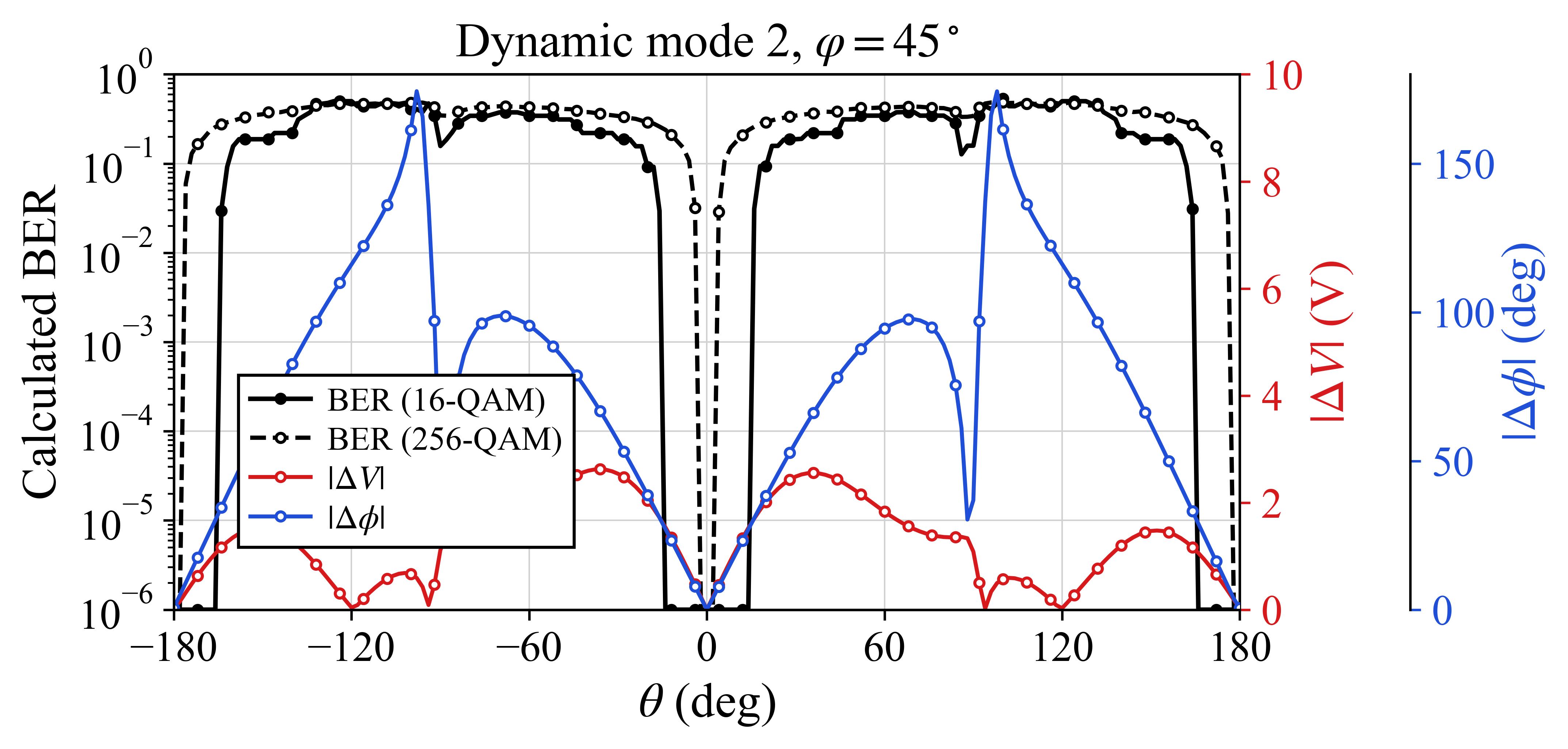}}\hfil\\
	\subfloat[]{\includegraphics[width=3.2in]{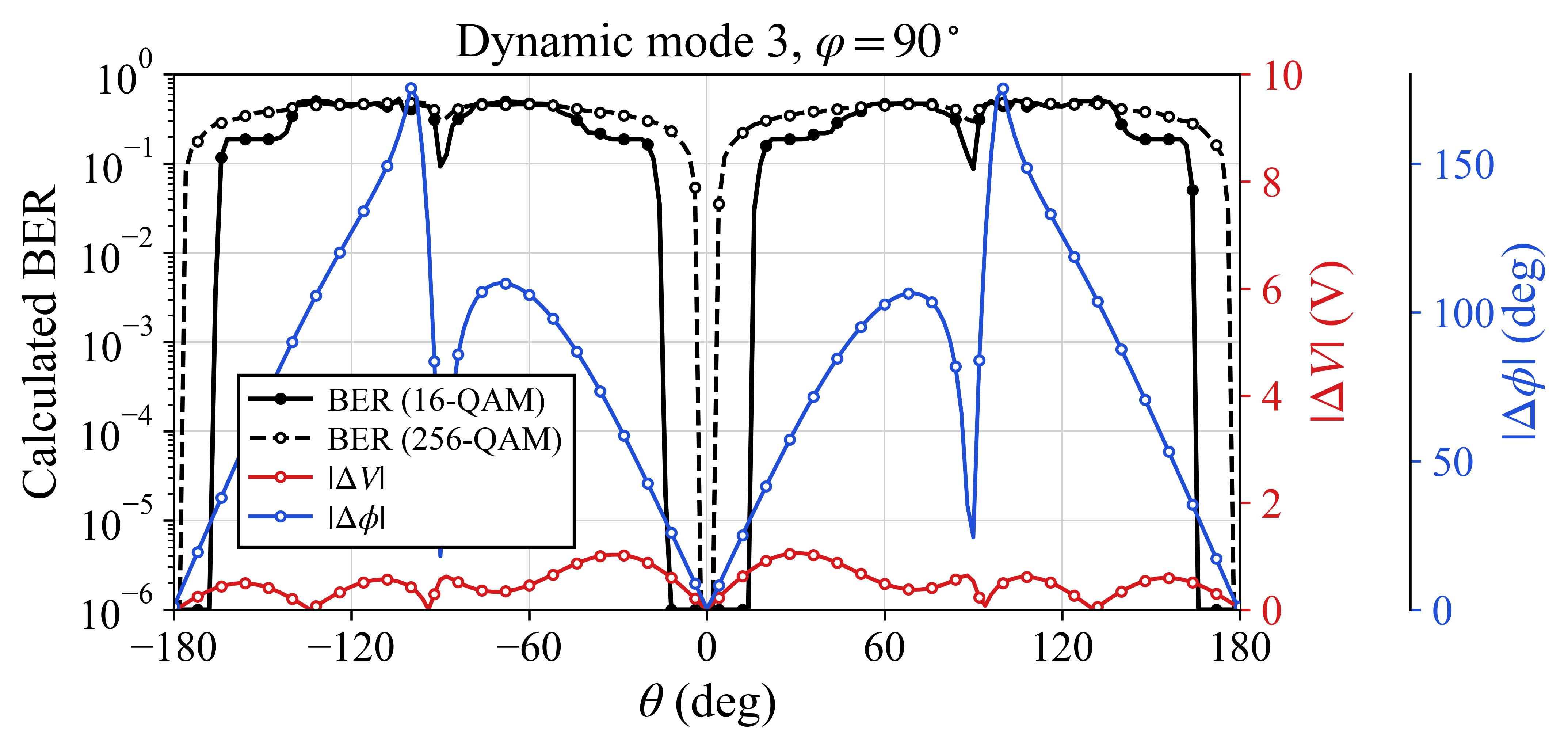}}\hfil
	\subfloat[]{\includegraphics[width=3.2in]{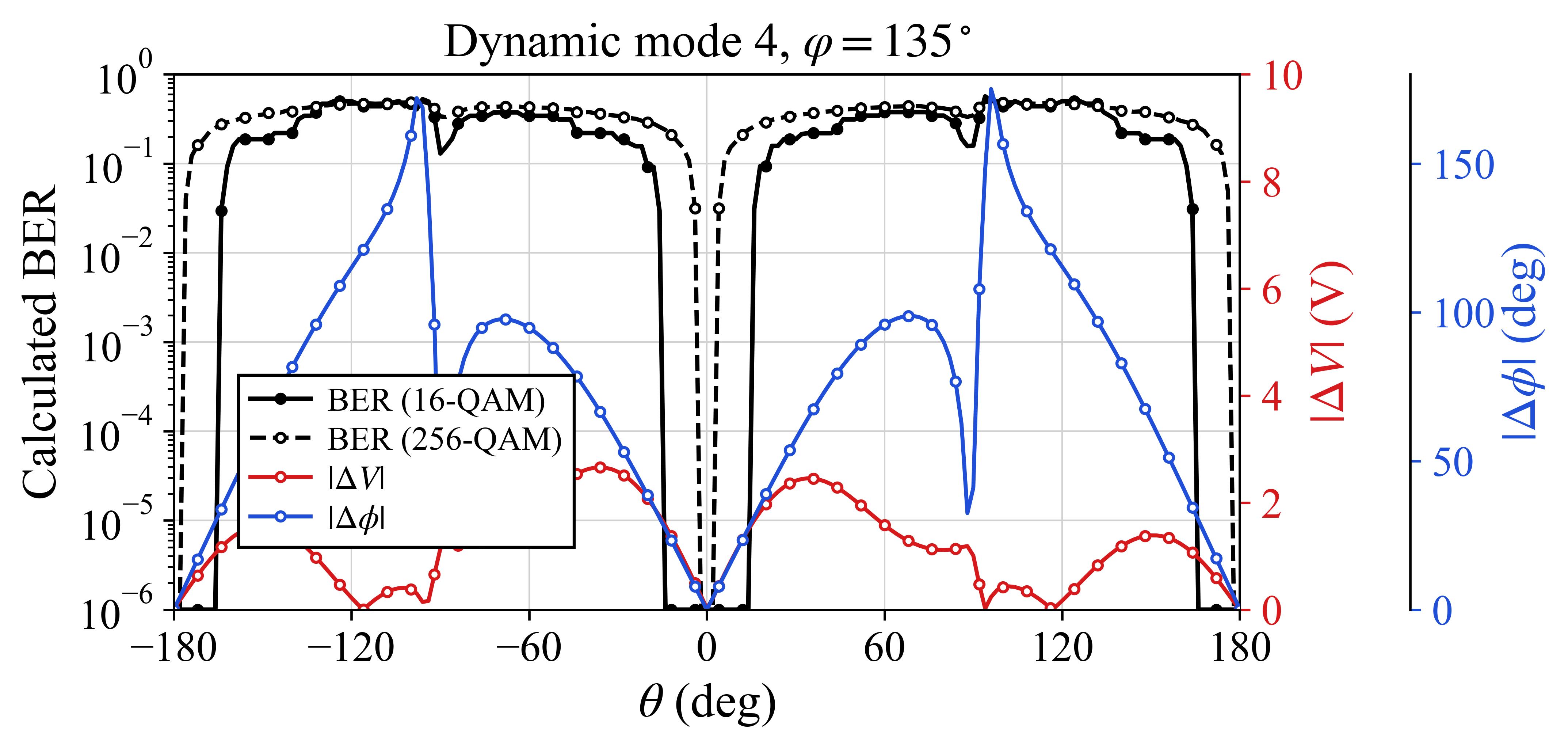}}\hfil
	\caption{Calculated bit error rate (BER), differential magnitude, and differential phase as functions of observation angle for the four dynamic switching modes: (a) $\varphi=0^\circ$, (b) $\varphi=45^\circ$, (c) $\varphi=90^\circ$, and (d) $\varphi=135^\circ$. Results are shown for both 16-QAM and 256-QAM modulations. The differential magnitude represents the amplitude of the information-bearing component, while the differential phase characterizes the angular variation of the phase pattern induced by differential port currents. Low-BER regions indicate angular sectors where the modulation constellation remains recoverable, defining the information beamwidth for each dynamic mode.}\label{fig:ber_patterns}
\end{figure*}
For the principal-axis switching modes ($\varphi=0^\circ$ and $90^\circ$) in Fig.~\ref{fig:dynamic_patterns}(a) and Fig.~\ref{fig:dynamic_patterns}(c), the odd-symmetric current distributions are oriented along the $x$ and $y$ directions, respectively, whereas the diagonal switching modes ($\varphi=45^\circ$ and $135^\circ$) in Fig.~\ref{fig:dynamic_patterns}(b) and Fig.~\ref{fig:dynamic_patterns}(d) produce odd-symmetric current patterns along the corresponding diagonal orientations. These differential current distributions directly reconfigure the aperture phase pattern, thereby determining the angular dependence of the information-bearing radiation. As a result, the direction of the information beam is governed by the spatial orientation of the differential current distribution, confirming the physical mechanism underlying the proposed reconfigurable directional modulation scheme.

\section{Design and Experiment in Wireless Communication}

\subsection{The Investigation of Information Beamwidth}
\begin{figure*}[!t]
	\centering
	\subfloat[]{\includegraphics[width=2.85in]{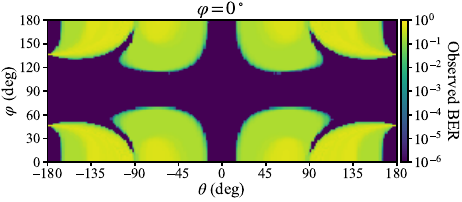}}\hfil
	\subfloat[]{\includegraphics[width=2.85in]{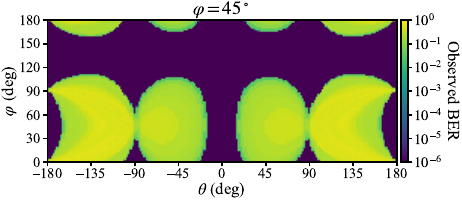}}\hfil\\[-1pt]
	\subfloat[]{\includegraphics[width=2.85in]{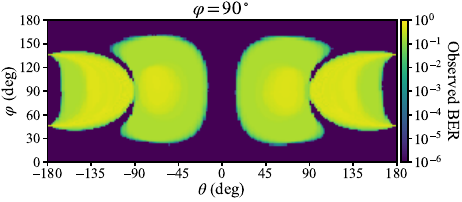}}\hfil
	\subfloat[]{\includegraphics[width=2.85in]{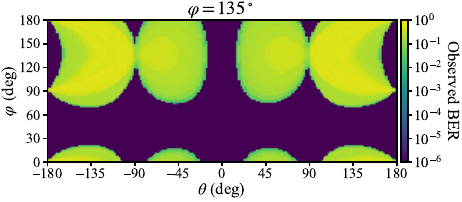}}\hfil\\[-1pt]
	\subfloat[]{\includegraphics[width=2.85in]{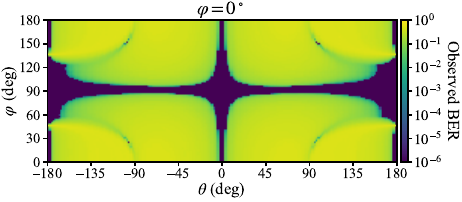}}\hfil
	\subfloat[]{\includegraphics[width=2.85in]{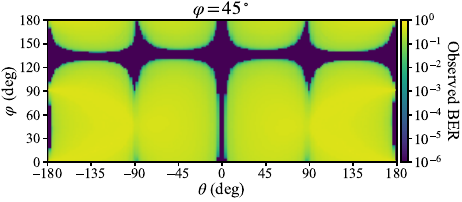}}\hfil\\[-1pt]
	\subfloat[]{\includegraphics[width=2.85in]{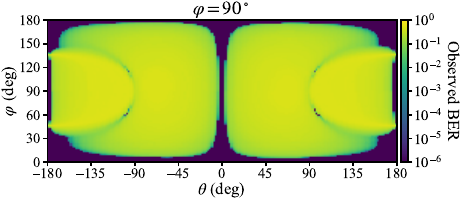}}\hfil
	\subfloat[]{\includegraphics[width=2.85in]{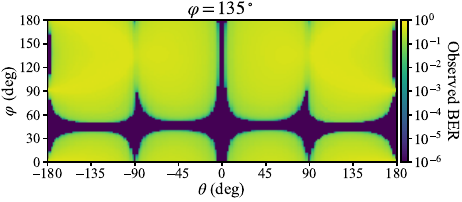}}\hfil
	\caption{Simulated all-angle observed BER distributions for the four reconfigurable information-beam modes. (a)--(d) 16-QAM results for dynamic mode~1 ($\varphi=0^\circ$), dynamic mode~2 ($\varphi=45^\circ$), dynamic mode~3 ($\varphi=90^\circ$), and dynamic mode~4 ($\varphi=135^\circ$), respectively. (e)--(h) Corresponding 256-QAM results for the same four dynamic modes. The observed BER color scale spans from $10^{-6}$ to $10^{0}$, and the observation angles follow the coordinate convention used in the array factor formulation.}
	\label{fig:all_angle_ber}
\end{figure*}

In this section, we study the communication performance of the designed dynamic omnidirectional meander antenna using MATLAB based on the communication channel model in~\cite{10286341}. A 48-kb pseudorandom bit sequence is modulated onto a 16-QAM signal using Gray coding, and the simulated amplitude and phase patterns of the two complementary states exported from HFSS are applied to the transmitted waveform. The SNR is set to 40~dB so that the calculated BER is dominated by directional modulation rather than by insufficient received power. We also evaluate the information beamwidth (IB), defined as the angular region over which BER $\leq 10^{-3}$, to quantify where the transmitted information can be recovered.

Fig.~\ref{fig:ber_patterns} plots the simulated BER together with the corresponding differential magnitude and phase for the four dynamic modes using 16-QAM and 256-QAM signals under complementary switching states with a 9-dB excitation power ratio. The differential quantities are extracted from the two complex far-field patterns of the complementary states: the differential magnitude is the absolute difference between the two state magnitudes, and the differential phase is the absolute unwrapped phase difference between the two states in the plotted cut. These quantities therefore describe the state-to-state variation of the far-field magnitude and phase patterns, rather than a separate radiation-power beam. The higher-order 256-QAM case produces a narrower recoverable angular region because the denser constellation has lower tolerance to the same switching-induced magnitude and phase errors. The angular dependence of the BER can be directly interpreted using the average--differential array factor decomposition derived in Section~II. Under dynamic switching, the time-averaged radiated power is given by $\bar{P}(\theta,\varphi) \propto |AF_{\mathrm{avg}}(\theta,\varphi)|^2 + |AF_{\Delta}(\theta,\varphi)|^2$, where the average term governs the intended radiation behavior, while the differential term captures the switching-induced amplitude and phase perturbations. The BER is therefore dominated by the relative contribution of the differential term in angular regions where switching-induced distortion outweighs the average radiation component.

\begin{figure*}[!t]
	\centering
	\begin{minipage}[t]{0.48\textwidth}
		\centering
		\includegraphics[width=\linewidth]{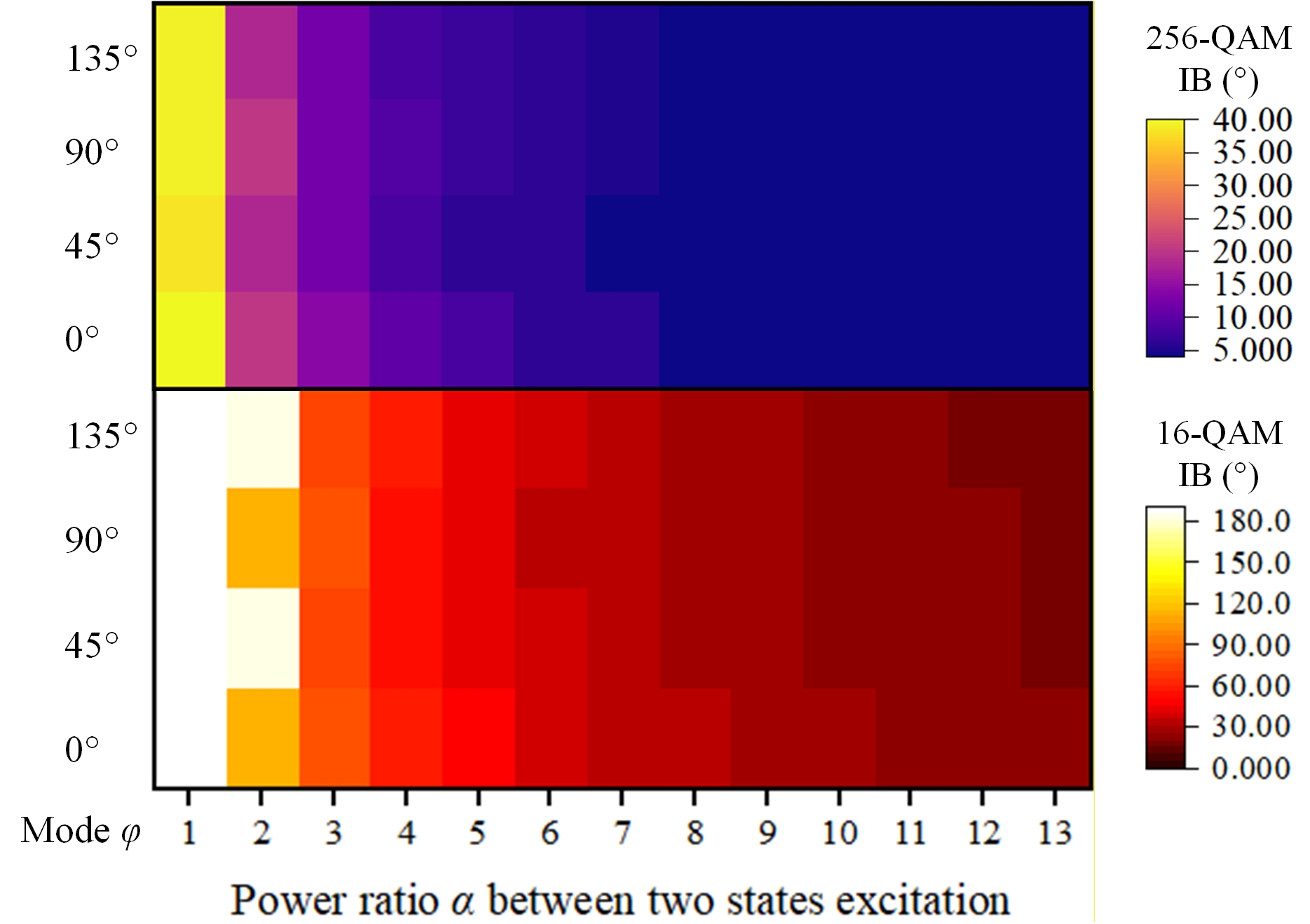}%
		\caption{Information beamwidth (IB) as a function of the excitation power ratio $\alpha$ between two switching states for different dynamic modes. The IB variation is primarily governed by the power ratio, which directly controls the differential aperture excitation and resulting phase pattern distribution. The use of 16-QAM and 256-QAM only affects the BER-based detection threshold used to extract IB, rather than the underlying beam-shaping mechanism.}
		\label{fig:ib_alpha}
	\end{minipage}\hfill
	\begin{minipage}[t]{0.48\textwidth}
		\centering
		\includegraphics[width=0.9\linewidth]{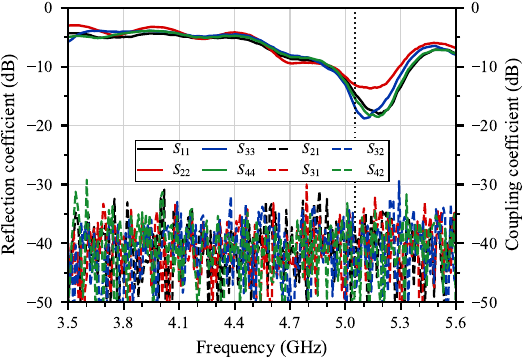}
		\caption{Measured $S$-parameter response of the fabricated four-port antenna. The solid traces show the reflection coefficients $S_{11}$, $S_{22}$, $S_{33}$, and $S_{44}$ on the left axis, while the dashed traces show the selected inter-port coupling coefficients $S_{21}$, $S_{31}$, $S_{32}$, and $S_{42}$ on the right axis.}
		\label{fig:measured_s11}
	\end{minipage}
\end{figure*}

\begin{figure*}[!b]
	\centering
	\includegraphics[width=0.84\textwidth]{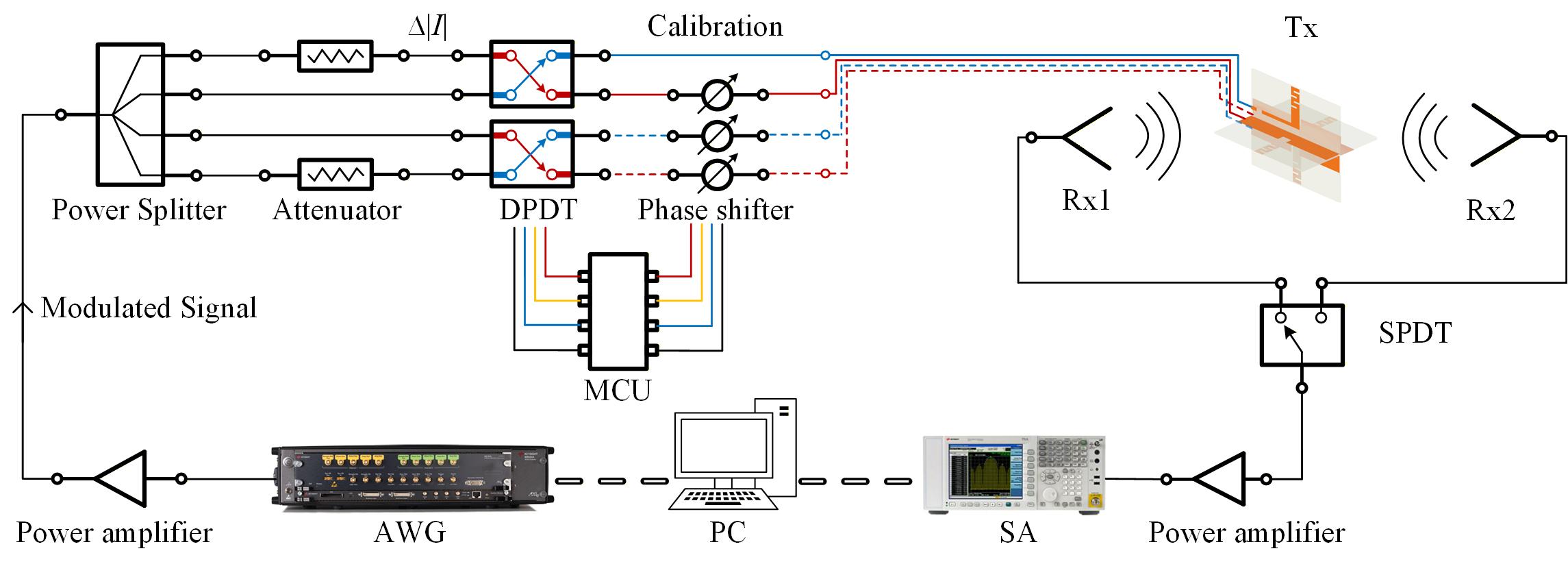}
	\caption{Block diagram of the measurement system used to evaluate the communication performance of the proposed cross-structured dynamic antenna.}
	\label{fig:measurement_schematic}
\end{figure*}

For a fixed power ratio of $\alpha=9$~dB, different switching configurations supported by the same antenna array lead to different angular distributions of the differential array factor $AF_{\Delta}(\theta,\varphi)$. Although the functional form of $\bar{P}(\theta,\varphi)$ remains unchanged, the spatial extent and magnitude of the differential term vary with the selected switching pair, producing distinct angular patterns of waveform distortion and BER.
The all-angle observed BER maps in Fig.~\ref{fig:all_angle_ber} extend the one-dimensional angular cuts by showing the complete spherical distribution of the recoverable-information region. The plotted quantity is the bit error rate obtained after demodulating the switched received waveform, so the color scale directly represents communication recoverability rather than only field magnitude or phase. For each dynamic mode, the E-plane cut exhibits a narrow low-BER angular sector, indicating that reliable demodulation is confined to a limited range of $\theta$ in the steering plane. The 256-QAM results show a narrower recoverable region than the 16-QAM results because the higher-order constellation is more sensitive to the same magnitude and phase distortion. In contrast, the corresponding H-plane cut remains low BER across the full plotted angular range from $\theta=-180^\circ$ to $180^\circ$. Thus, the proposed antenna does not create a narrow power beam in all directions; instead, it combines direction-selective information recovery in the selected E-plane cut with omnidirectional information recovery in the corresponding H-plane cut. This behavior is consistent with the intended omnidirectional communication function of the antenna, while the switching-induced differential aperture response provides angular selectivity in the E-plane information beam.

The azimuthal location of this full-$\theta$ low-BER cut follows the selected dynamic mode. As the switching state is changed from $\varphi=0^\circ$ to $45^\circ$, $90^\circ$, and $135^\circ$, the low-BER band appears at the corresponding azimuthal angle in Fig.~\ref{fig:all_angle_ber}. This one-to-one correspondence between the selected switching mode and the azimuthal location of the recoverable H-plane confirms that the information beam has been successfully rotated by reorienting the differential aperture component $AF_{\Delta}$. Away from the selected plane, the simulated BER rapidly approaches the high-error region of the color scale because the differential component introduces angle-dependent magnitude and phase perturbations that corrupt the 16-QAM symbols.

Fig.~\ref{fig:ib_alpha} demonstrates that the information beamwidth scales with the excitation power ratio $\alpha$ due to the resulting differential current imbalance and associated phase redistribution across the antenna aperture. The same IB compression trend is maintained for all dynamic modes, confirming that the beamwidth control mechanism is invariant under information-beam rotation. The modulation schemes serve only as BER evaluation references and do not alter the fundamental aperture-driven IB formation.

\subsection{Experimental Communication Performance}

\begin{figure*}[!t]
	\centering
	\subfloat[]{%
		\includegraphics[width=3.2in]{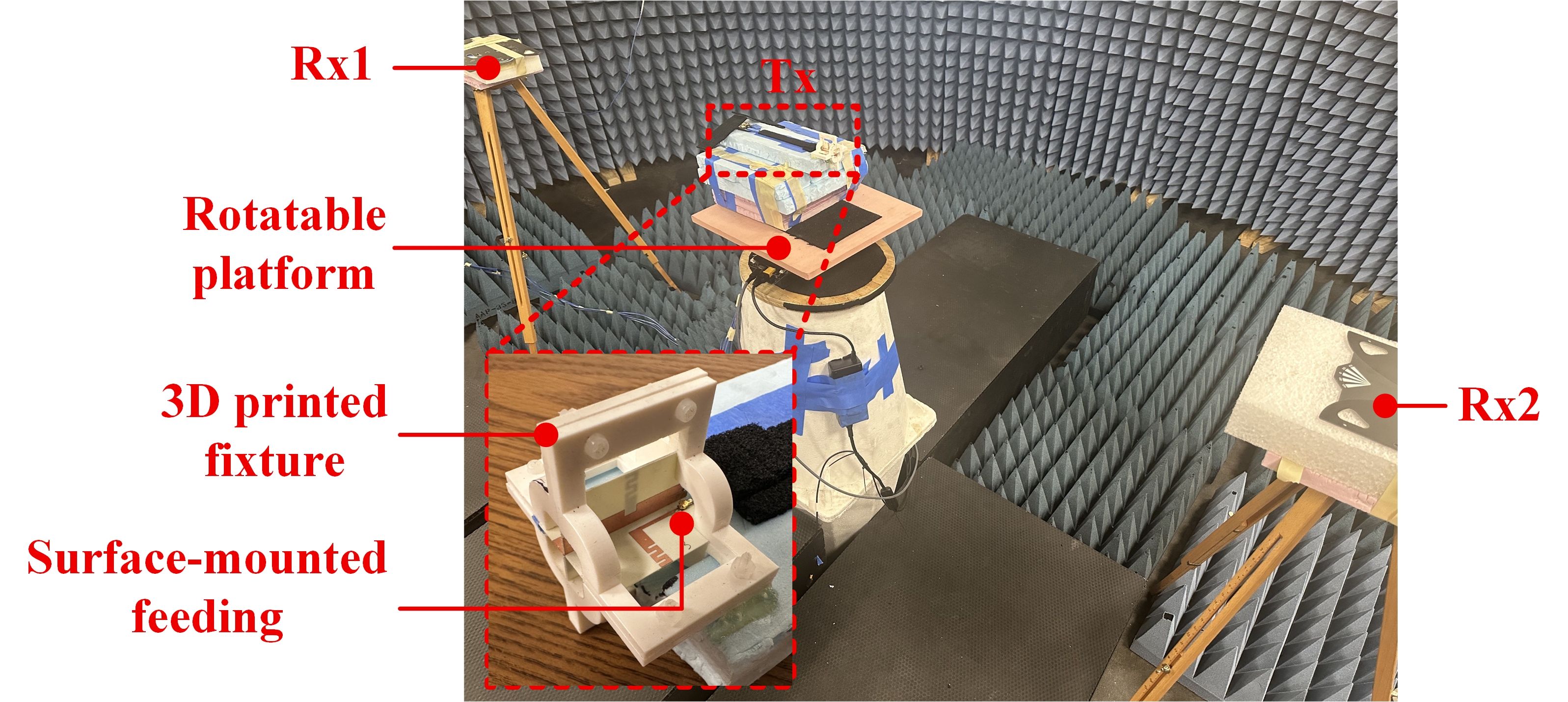}
	}\hfil
	\subfloat[]{%
		\includegraphics[width=3.6in]{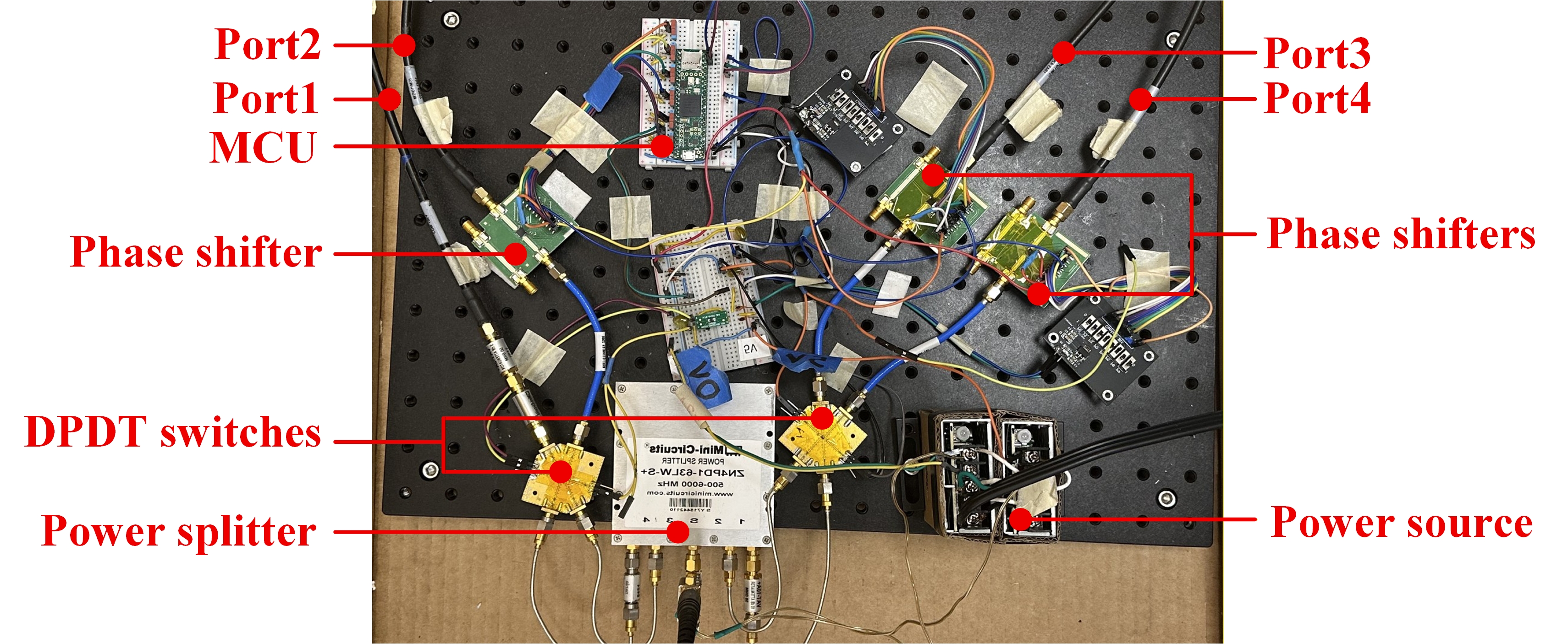}
	}
	
	\caption{Measurement setup for evaluating the communication performance of the proposed cross-structured dynamic antenna: (a) experimental measurement setup, and (b) switching system configuration.}
	\label{fig:measurement_setup}
	\vspace{3pt}
	
	\subfloat[]{\includegraphics[width=2.85in]{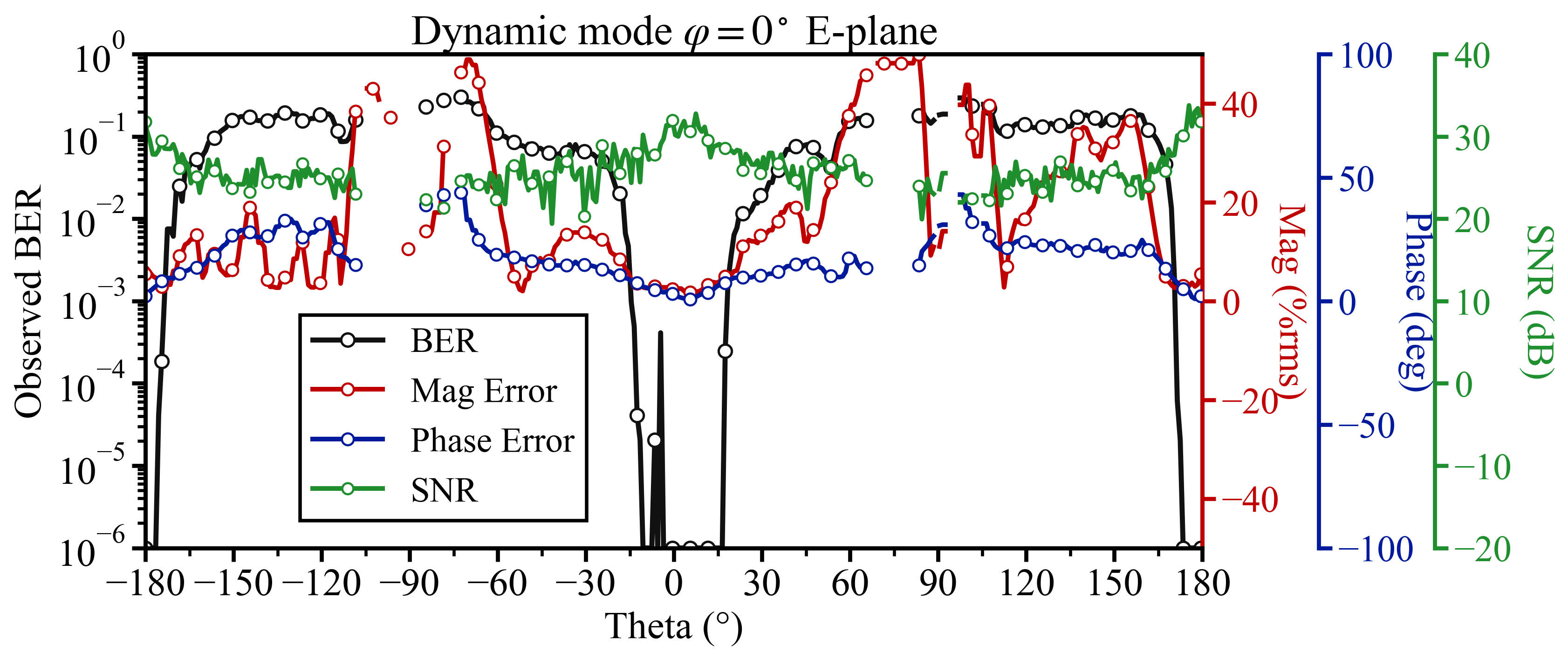}}\hfil
	\subfloat[]{\includegraphics[width=2.85in]{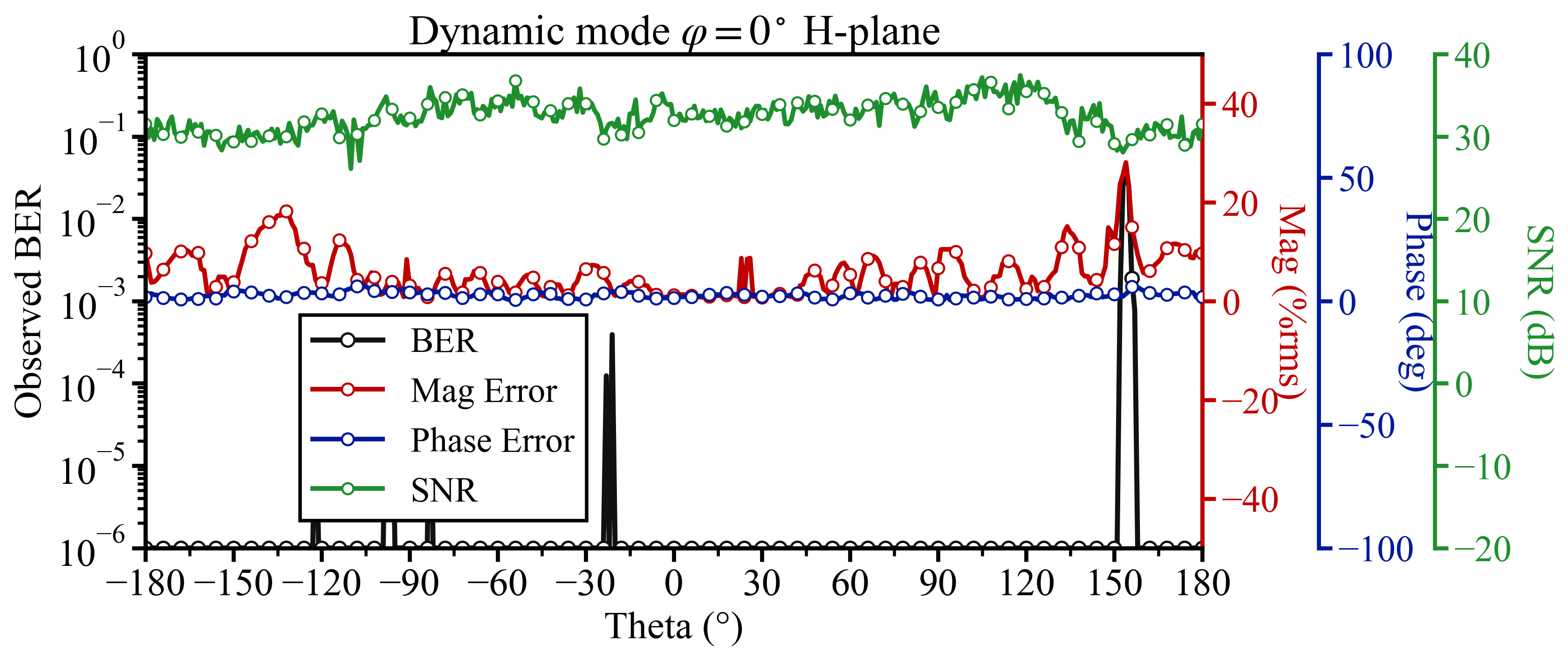}}\hfil\\[-2pt]
	\subfloat[]{\includegraphics[width=2.85in]{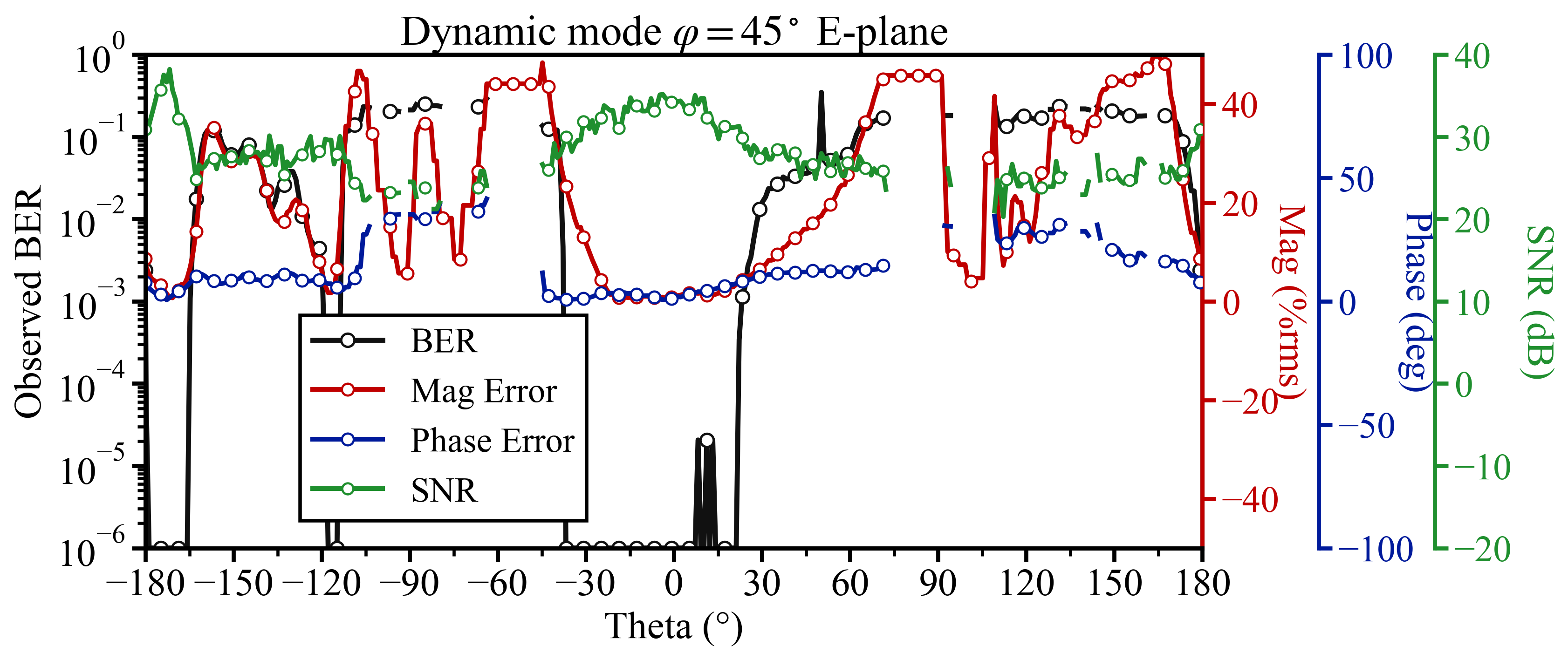}}\hfil
	\subfloat[]{\includegraphics[width=2.85in]{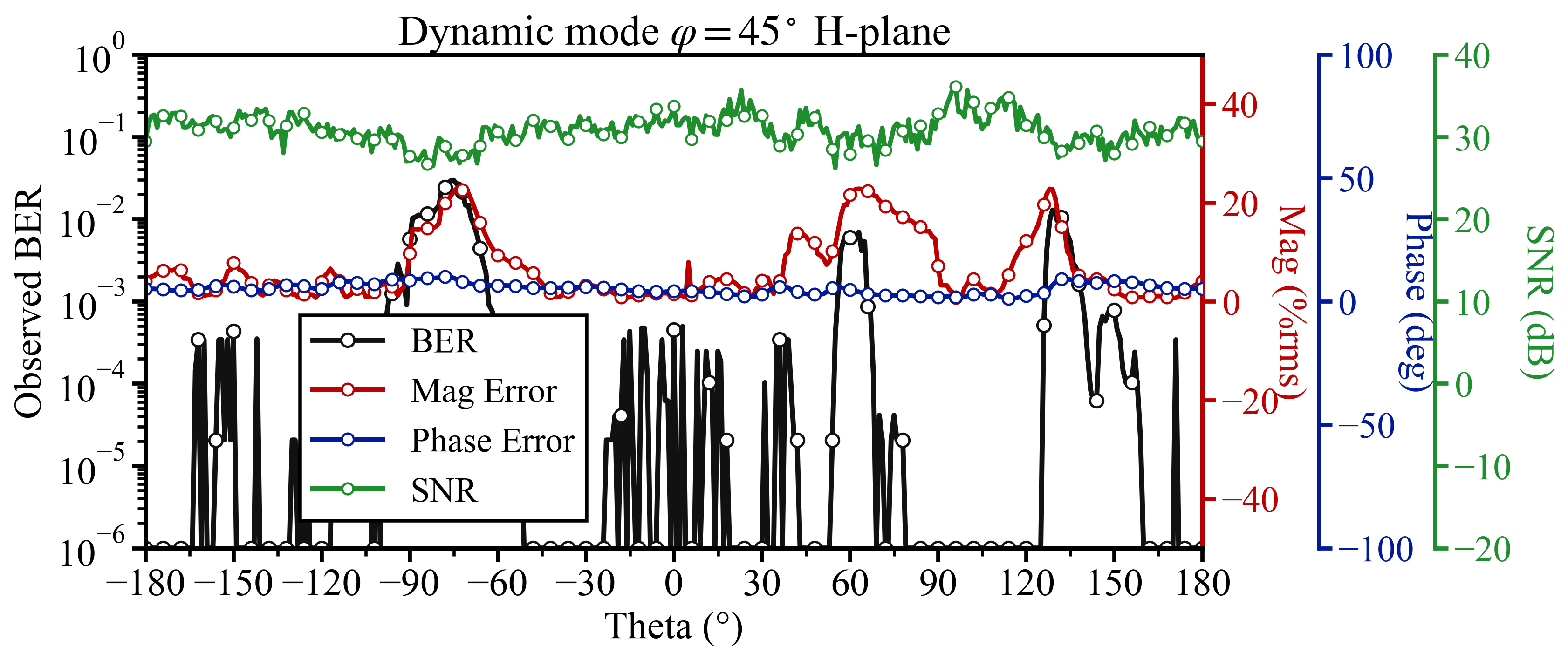}}\hfil\\[-2pt]
	\subfloat[]{\includegraphics[width=2.85in]{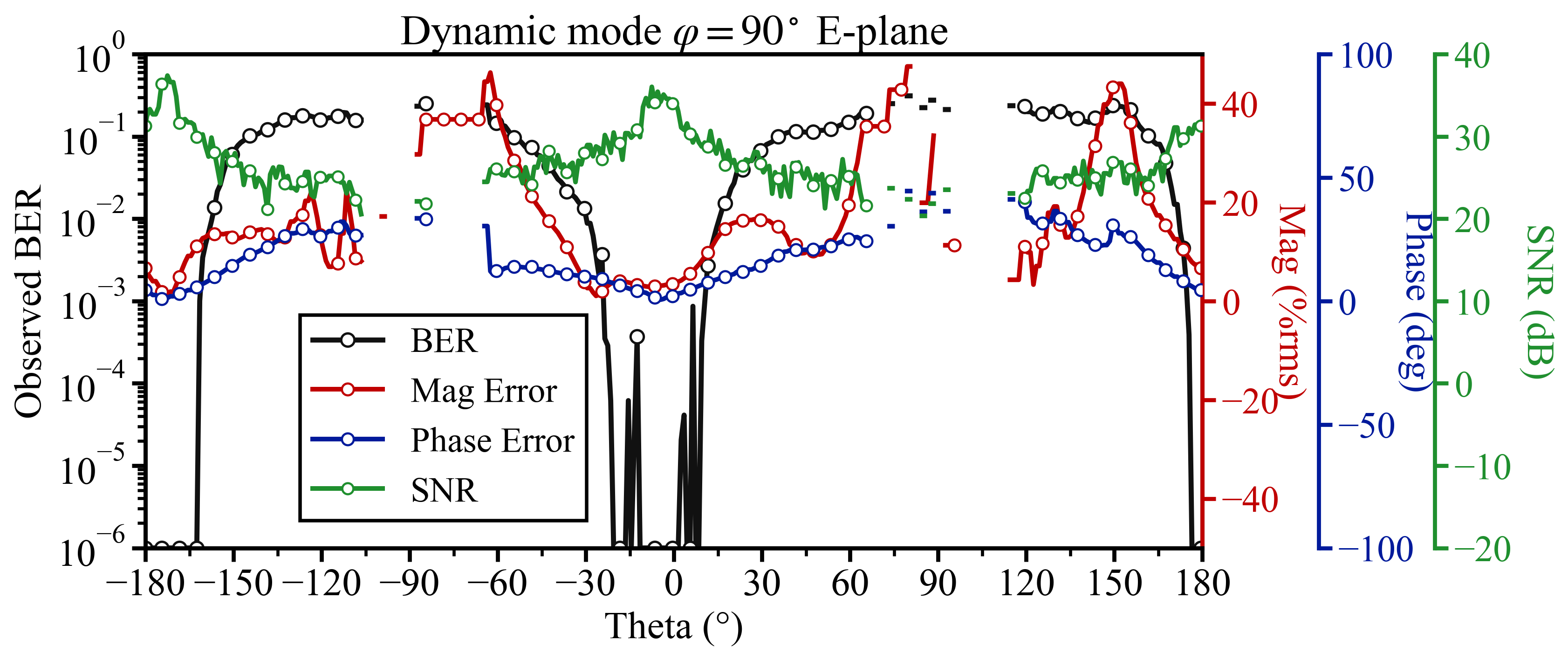}}\hfil
	\subfloat[]{\includegraphics[width=2.85in]{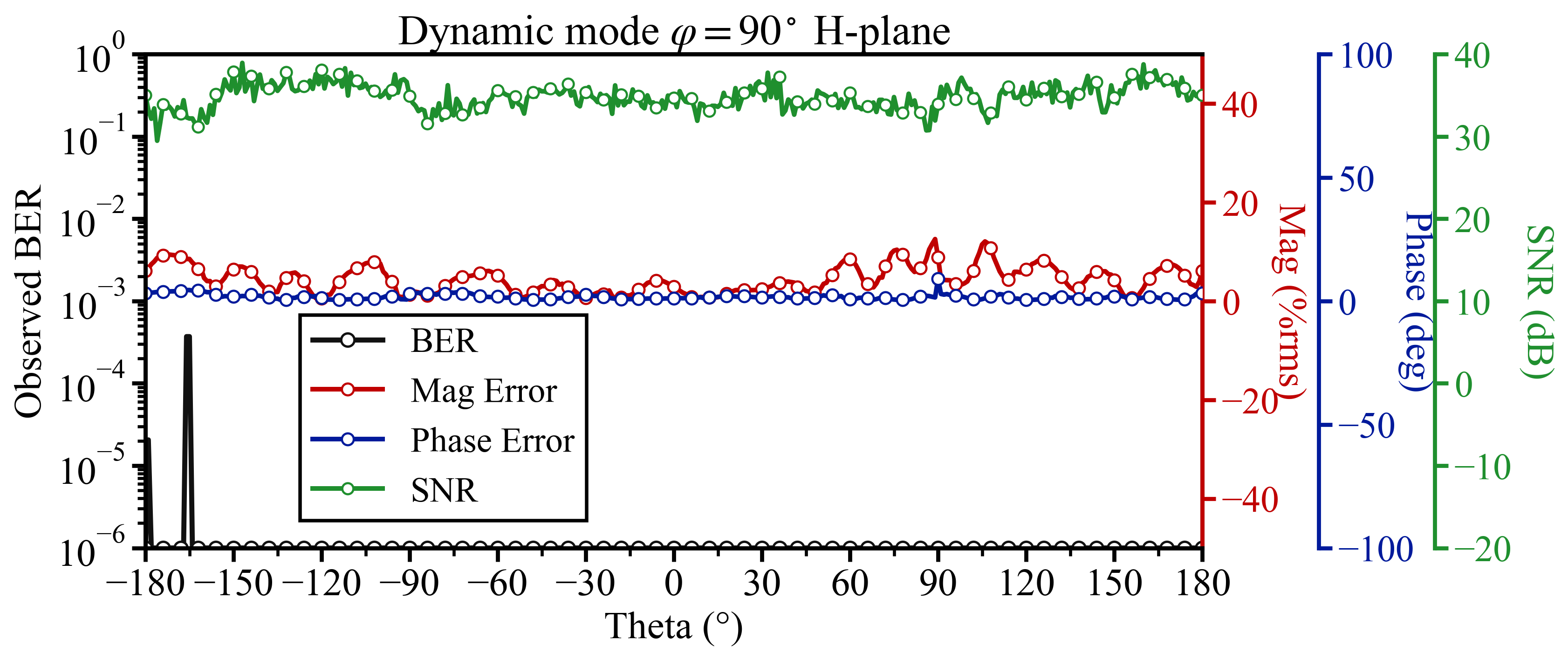}}\hfil\\[-2pt]
	\subfloat[]{\includegraphics[width=2.85in]{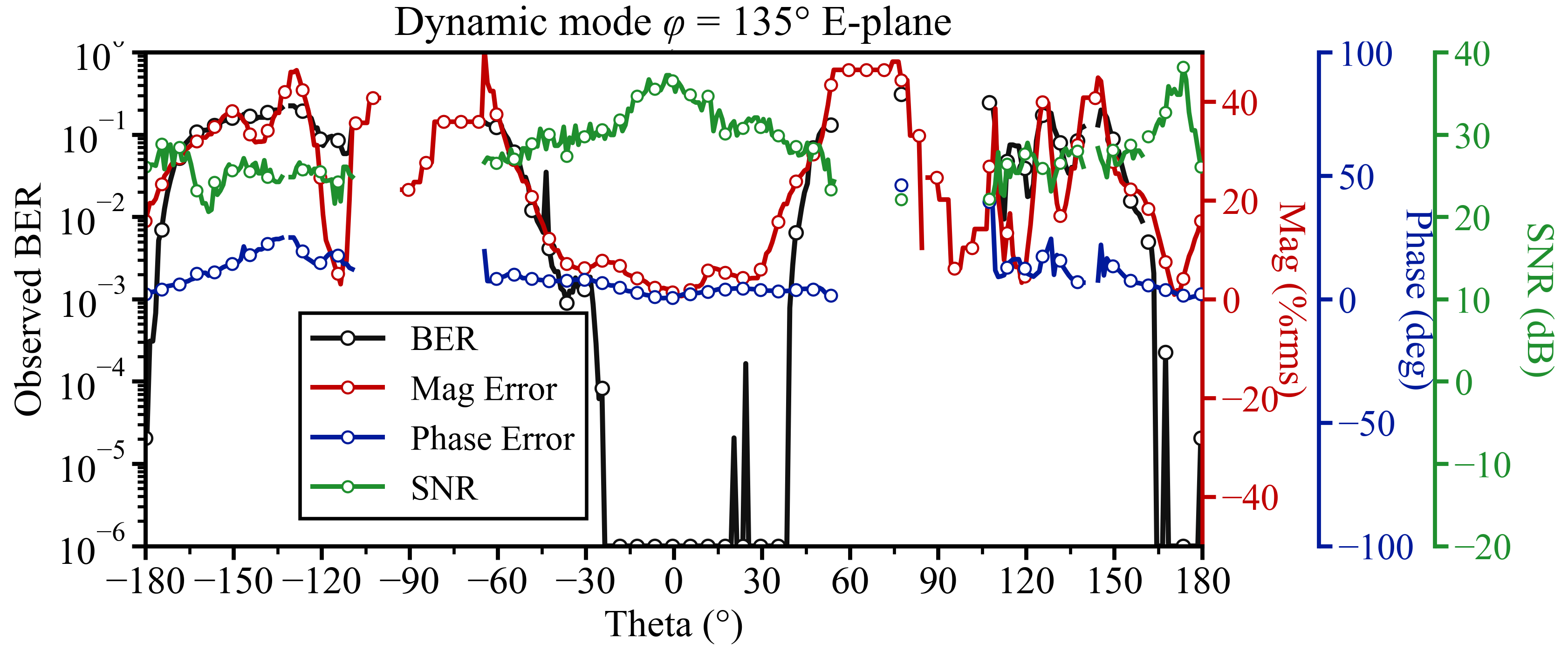}}\hfil
	\subfloat[]{\includegraphics[width=2.85in]{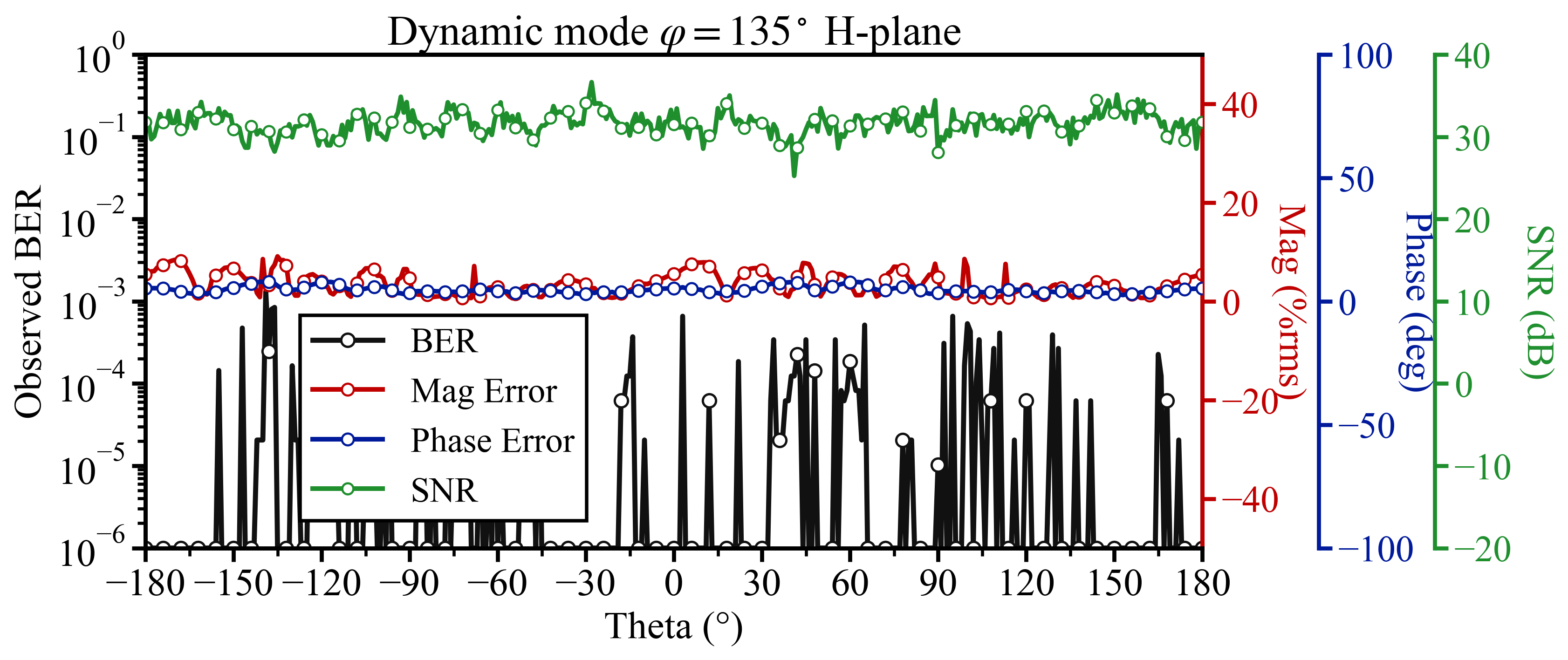}}\hfil
	\caption{Measured communication performance of the proposed cross-structured dynamic array at 5.05~GHz using 16-QAM. The observed BER, magnitude error, phase error, and received SNR versus $\theta$ are presented for the E- and H-plane measurements of four switching modes: (a) and (b) switching between antenna ports 1 and 3, corresponding to $\varphi=0^\circ$; (c) and (d) switching between port groups 12 and 34, corresponding to $\varphi=45^\circ$; (e) and (f) switching between antenna ports 2 and 4, corresponding to $\varphi=90^\circ$; and (g) and (h) switching between port groups 13 and 24, corresponding to $\varphi=135^\circ$.}
	\label{fig:measured_cross_dynamic_ber}
\end{figure*}

The communication measurement architecture is summarized in Fig.~\ref{fig:measurement_schematic}. A digitally modulated 16-QAM signal at 5.05~GHz was generated by the arbitrary waveform generator (AWG), amplified, and then divided into four RF branches by a broadband power splitter. The branch amplitudes were adjusted by attenuators to realize the calibrated switching power ratio used in the experiment, while the phase shifters were used for branch-dependent phase calibration introduced by cables, switches, and RF components. Two DPDT switches, driven by a microcontroller unit (MCU), selected the complementary excitation states associated with each dynamic mode, so that the same fabricated antenna could be measured under the four port-group configurations listed in Table~\ref{tab:table2}.

The fabricated dynamic array was used as the transmitter, as shown in Fig.~\ref{fig:measurement_setup}. Two broadband receiving antennas were arranged to capture the front and back angular responses, and their outputs were selected through an SPDT switch before being sent to the signal analyzer (SA). During the angular sweep, the received waveform was demodulated to extract the observed BER, magnitude error, phase error, and received SNR at each observation angle. This setup allows the measured BER variation to be attributed to the intended switching-induced differential magnitude and phase response of the antenna, rather than simply to received-power variation.

The measured $S$-parameter response in Fig.~\ref{fig:measured_s11} verifies that the fabricated four-port antenna provides an appropriate RF platform for the communication experiment at 5.05~GHz. The four reflection coefficients remain reasonably matched around the operating frequency, while the selected inter-port coupling traces are substantially lower than the reflection responses over the same band. This separation indicates that the measured communication behavior is governed primarily by the intended switching-induced aperture modulation rather than by excessive port coupling. Therefore, the BER variation reported below can be interpreted as the effect of the designed differential magnitude and phase response, rather than as a consequence of poor matching or unintended coupling between antenna ports.
The fabricated prototype and measurement configuration are shown in Fig.~\ref{fig:measurement_setup}, where the final array was fixed in a 3D printed dielectric holder. The required phase inversion was implemented using the additional electrical length introduced by two SMA adapters. The four antenna ports are connected to the external switching network, which implements the two complementary excitation states listed in Table~\ref{tab:table2}. Unlike conventional pattern-reconfigurable or phased-array DM implementations, the proposed antenna does not steer a high-gain radiation beam. Instead, it reconfigures the odd-symmetric differential aperture component $AF_{\Delta}$ while maintaining a nearly omnidirectional average radiation response. This distinction is important because the information beam is formed by the angular distribution of recoverable modulation, rather than by the maximum of the radiated power pattern.

The measured communication results in Fig.~\ref{fig:measured_cross_dynamic_ber} experimentally verify the central novelty of this work: a compact four-port cross-structured antenna can rotate the information-recoverable region among the four designed dynamic modes using only antenna-level switching. Using the BER $\leq 10^{-3}$ criterion, the E-plane measurements show that reliable demodulation is concentrated around the intended information-beam sector for each mode, with the dominant low-BER regions located near the centered beam direction and the corresponding angular counterpart introduced by the spherical scan coordinate. Outside these regions, the BER increases by orders of magnitude and is accompanied by large magnitude and phase errors; in several off-beam angular ranges, the vector signal analyzer cannot recover a stable constellation, which is why no valid BER points are reported. This behavior is observed while the received SNR remains high, above 19.4~dB in all valid E-plane samples, confirming that the BER degradation is not caused by insufficient received power but by the designed directional waveform distortion. The H-plane results provide complementary evidence: for each measured mode, the BER remains below $10^{-3}$ over nearly the full measured angular range, with substantially smaller magnitude and phase variations and SNR above 25.3~dB in all valid H-plane samples. Therefore, the experiment confirms the intended operating behavior: dynamic switching creates a reconfigurable and spatially confined information beam in the selected E-plane cut, while maintaining omnidirectional low-BER information recovery in the corresponding H-plane cut.

The four E-plane measurements in Fig.~\ref{fig:measured_cross_dynamic_ber}(a), (c), (e), and (g) further show that the low-BER sector follows the selected differential aperture orientation associated with the 1--3, 12--34, 2--4, and 13--24 port groups. This behavior agrees with the simulated all-angle BER maps in Fig.~\ref{fig:all_angle_ber}, where the recoverable-information region rotates with the selected dynamic mode. The simultaneous BER, magnitude-error, phase-error, and SNR traces also clarify the degradation mechanism: high-BER E-plane samples generally coincide with increased magnitude and phase errors, while the SNR does not collapse in the same manner. Therefore, the measured communication failure is mainly caused by switching-induced waveform distortion predicted by the average--differential array factor model, not by a simple received-power null.

\section{Conclusion}
This paper demonstrated a compact cross-structured dynamic antenna for planar physical-layer security based on antenna-level directional modulation. Experimental results using 16-QAM under high-SNR conditions show that reliable communication is confined to the intended information-beam sector in the selected E-plane cut for each dynamic mode, while the BER degrades significantly outside that sector, confirming strongly angle-dependent communication performance. The observed behavior is enabled by switching-induced differential magnitude and phase modulation within a four-element planar array of meander-line monopole antennas. In contrast, the corresponding H-plane cuts remain quasi-static and maintain omnidirectional low-BER information recovery, which is consistent with the intended omnidirectional communication behavior of the antenna. An average--differential array factor formulation explains how different switching pairs rotate the recoverable-information region through their impact on the angular distribution of the differential aperture component. The antenna is realized on a single-layer commercial substrate with a compact footprint and validated using a low-complexity four-path switching network composed of commercial RF components. These results demonstrate a practical and hardware-efficient dynamic array architecture for secure omnidirectional communication and suggest its applicability as a building block for future distributed and reconfigurable antenna systems.

\raggedbottom

\end{document}